\documentclass[12pt]{article}

\usepackage{graphicx}
\usepackage{amsmath}
\usepackage{amssymb}

\usepackage{caption}
\usepackage{titlesec}
\titleformat{\section}
  {\normalfont\fontsize{15}{18}\bfseries}{\thesection}{1em}{}
\begin{document}

${}$\\
\begin{center}
\vspace{36pt}
{\large \bf De Sitter Universe from Causal Dynamical Triangulations \\ \vspace{12pt} 
without Preferred Foliation}
\vspace{48pt}

{\sl S. Jordan }
and {\sl R. Loll}

\vspace{24pt}

{\footnotesize

Radboud University Nijmegen, \\
Institute for Mathematics, Astrophysics and Particle Physics, \\
Heyendaalseweg 135, NL-6525 AJ Nijmegen, The Netherlands.\\ \vspace{8pt} 
{email: s.jordan@science.ru.nl, r.loll@science.ru.nl}\\

}

\vspace{48pt}

\end{center}

\begin{center}
{\bf Abstract}
\end{center}
We present a detailed analysis of 
a recently introduced version of Causal Dyna\-mi\-cal Triangulations (CDT) that does not rely on a distinguished time slicing.
Focussing on the case of 2+1 spacetime dimensions, we analyze its geometric and causal properties,
present details of the numerical set-up and explain how to extract ``volume profiles". Extensive Monte Carlo measurements
of the system show the emergence of a de Sitter universe on large scales from the underlying quantum ensemble, similar
to what was observed previously in standard CDT quantum gravity. This provides evidence that the distinguished time
slicing of the latter is not an essential part of its kinematical set-up. 

\newpage

\section{Introduction}

First attempts in the 1990s to define quantum gravity nonperturbatively with the help of Dynamical Triangulations (DT)
were based on an intrinsically Euclidean path integral, whose configuration space consists of 
four-dimensional, curved Riemannian spaces. DT works with a regularized version of this space in terms of triangulations,
piecewise flat spaces of positive definite metric signature. Its elementary building block is a four-simplex, 
a generalization to four dimensions of a triangle (two-simplex) and a tetrahedron (three-simplex). An individual building 
block is a piece of flat four-dimensional Euclidean space and therefore does not carry any curvature. 
However, numerical investigations of the 
nonperturbative dynamics of DT quantum gravity found that it has neither a large-scale limit 
compatible with general relativity \cite{Ambjorn:1991pq,Agishtein:1992xx,Catterall:1994pg}, nor a second-order phase
transition allowing for a continuum limit in the sense of lattice quantum field theory \cite{Bialas:1996wu, deBakker:1996zx}.
Despite the negative nature of this result, the fact that the model contains criteria which could be used for
its falsification should be appreciated. For many other candidate theories of quantum gravity this is not obviously the case.

Causal Dynamical Triangulations (CDT) were introduced in \cite{al} in an attempt to overcome these problems. 
The key new idea of CDT is to incorporate aspects of the {\it causal structure} of classical general relativity 
at a more fundamental level into the nonperturbative gravitational path integral.\footnote{By contrast, Euclidean gravity does not 
distinguish between space and time, and therefore has no causal structure and no notion of causality.} The elementary
building blocks of CDT quantum gravity are flat four-simplices of {\it Lorentzian} signature, that is, pieces of
Minkowski space. The carrier space of the corresponding path integral consists of piecewise flat simplicial
manifolds assembled from these building blocks. In addition, each path integral history has a distinguished discrete
fo\-liation and an associated notion of (discrete) proper time, which ensures the pre\-sence of a well-defined causal structure globally
(see \cite{cdtreviews,physrep} for details on motivation, construction and results in CDT).
Numerical simulations in 3+1 dimensions have shown that these modifications lead to a completely different quantum dynamics,
compared to the earlier Euclidean DT model: CDT quantum gravity in 3+1 dimensions contains a phase whose nonperturbative
ground state of geometry is extended, macroscopically four-dimensional and on large scales can be matched to a de
Sitter universe \cite{cdt4d,desitter}. Moreover, the theory has recently been shown to possess a second-order phase transition, 
which according to standard arguments is a prerequisite for the existence of a continuum limit \cite{trans}.

The causal structure of CDT is realized by putting together its simplicial building blocks such that each 
CDT configuration has a product structure, not just at the level of
topology -- usually chosen as $[0,1]\times{}^{(3)}\Sigma$, for fixed three-topology ${}^{(3)}\Sigma$ -- but {\it as triangulations}.
A (3+1)-dimensional CDT geometry consists of a sequence of slabs or layers, each of thickness 1, which may be thought of
as a single unit of proper time. The orientation of the light cones of all four-simplices in a given slab is consistent 
with this notion of time. In this way the causal structure becomes linked to a preferred discrete foliation of spacetime.

To understand better how the preferred time foliation on the one hand and the causal structure on the other
contribute to the evidence of a good classical limit -- the key 
distinguishing feature of CDT quantum gravity -- it would be highly desirable to disentangle these
two elements of ``background structure". We recently proposed a modification of standard CDT which does exactly that
\cite{jl}. The main idea, applicable in any spacetime dimension $d$, is to enlarge the set of $d$-simplices by new types of 
simplicial building blocks, also pieces of $d$-dimensional Minkowski space, but with different
link type assignments and therefore a different orientation of the light cone relative to the boundaries of the simplex. 
Including the new building blocks is in general not compatible with the preferred foliation of CDT. Nevertheless, 
with a suitable choice of gluing 
rules one can still obtain Lorentzian simplicial manifolds with a well-defined causal structure, at least locally. 
In this way the issue
of causality becomes dissociated from the notion of a preferred foliation. The interesting question is then
whether the quantum-gravitational model using ``nonfoliated CDT" can reproduce the results of the
standard formulation, especially those concerning the large-scale properties of quantum spacetime. The main conclusion
of the present paper, previously announced in \cite{jl}, is that it {\it can}, at least in space-time dimension 2+1.
At least in higher dimensions, this result appears to weaken the potential link of CDT quantum gravity with 
Ho\v rava-Lifshitz gravity \cite{hl,phasediagram},
where the presence of a preferred foliation is a key ingredient.\footnote{This differs
from 1+1 dimensions, where CDT is a quantum-mechanical system of a single length variable and has been shown 
to coincide with projectable Ho\v rava-Lifshitz gravity \cite{2dcdthl}.}
 
To summarize, our intention is to get rid of the {\it distinguished} foliation (and associated discrete time label $t$) of CDT, 
whose leaves for integer $t$ coincide with simplicial spatial hypermanifolds consisting entirely of {\it space}like
subsimplices of codimension 1. This does not mean that the nonfoliated CDT configurations cannot in principle be foliated
with respect to some continuous notion of time, but only that there is in general no canonical way of doing this in terms of
a distinguished substructure of the triangulated spacetimes.

Of course, having {\it a} notion of time at the level of the regularized geometries is important. In the standard formulation
of CDT, we get a notion of (discrete proper) time $t$ for free, simply by counting consecutive ``slabs", as described earlier. 
The presence of this time label allows us to define a transfer matrix and an associated propagator with the correct 
behaviour under composition (whose continuum limit in dimension 1+1 can be computed analytically \cite{al}), 
and prove a version of reflection positivity \cite{3d4d}. 
Yet another advantage of having an explicit time variable is that we can construct observables like the volume profile, 
which measures the distribution of spatial volume as a function of time. The analysis of these volume profiles in 
3+1 dimensions has been crucial in relating the large-scale behaviour of CDT to a de Sitter cosmology
in the continuum \cite{desitter,semiclassical}.\footnote{One does not know a priori whether $t$ or any other
time label of the regularized model will assume a distinguished physical role in the continuum theory; this can only
be determined by studying suitable continuum {\it observables}, like the volume profiles.}

By contrast, in the enlarged CDT set-up we will be considering, typical triangulations will be more complicated, 
in the sense that the purely spatial subsimplices of codimension 1 will no longer align themselves into a neat sequence 
of simplicial hyper{\it manifolds}, but instead will form branching structures, as will be explained in more detail
below. This also implies that we no longer have a distinguished time variable 
at our disposal. 
Nevertheless, as we shall demonstrate for the nontrivial case of 2+1 dimensions, it is possible to construct a meaningful 
time variable, whose restriction to
standard, nonbranching CDT configurations agrees with the usual proper time label $t$. This will enable us to
extract volume profiles from the numerical simulations and compare their dynamics with that of
the standard formulation.

There are a number of good reasons for beginning our investigation in 2+1 dimensions, as we 
are doing in the present work. In standard CDT, the 
large-scale properties of the quantum universe generated by the nonperturbative
quantum dynamics are qualitatively very similar to those in 3+1 dimensions, and are well described by a (three-dimensional)
Euclidean de Sitter universe \cite{3dcdt,benedettihenson}. 
Furthermore, as we will describe in Sec.\ \ref{sec:numsetup}, the nonfoliated CDT model requires 
new Monte Carlo moves, which are significantly more difficult to implement than the generalized Pachner moves used 
in standard CDT. Writing the simulation software for the new model becomes a very challenging task already in 2+1 dimensions. 
At the same time, the increased complexity of the simulation software leads to longer simulation times. 
It is not even clear currently whether analogous simulations in 3+1 dimensions could be performed with acceptable 
running times using contemporary simulation hardware.

Before embarking on our exploration of the dynamics of nonfoliated CDT, let us comment briefly on prior work which
considered explicitly a possible relaxation of CDT's strict time slicing.\footnote{In CDT in 1+1 dimensions, one can
rederive the exact continuum propagator without in\-voking an explicit proper-time slicing (see, for example, \cite{durhuuslee}).
It is unclear how to generalize such a construction to higher dimensions.}
A soft way of relaxing the foliation in 1+1 dimensions was studied in \cite{markopoulousmolin}, where under
certain conditions the timelike links 
were allowed to have varying length. In this approach the foliation is still present, 
since the connectivity of the underlying triangulation is unchanged, but the individual leaves of the foliation are 
not placed at equidistant intervals. The authors argued that this should not affect the continuum limit of the model.
A similar idea in 2+1 dimensions was considered in \cite{konopka}, where it was also suggested to add new
elementary building blocks to CDT, which extend over two slabs of the foliation instead of one. This study
did not include details of how the path integral of the corresponding generalization of CDT quantum gravity should
be formulated or simulated. 

The remainder of this article is structured as follows.
We begin with a brief review of Causal Dynamical Triangulations in Sec.\ \ref{cdt}. In Sec.\ \ref{sec:ncdt_2d} we discuss 
aspects of CDT without preferred foliation in 1+1 dimensions, to illustrate the basic geometric idea behind the enlarged model. 
Sec.\ \ref{sec:nfct21} contains a detailed study of the kinematical aspects of nonfoliated CDT  
in 2+1 dimensions. Sec.\ \ref{sec:actions} deals with actions and the Wick rotation, and 
Sec.\ \ref{sec:numsetup} summarizes the new numerical set-up. This includes
an overview of the Monte Carlo moves in 2+1 dimensions, a prescription for how a notion of time can be
defined on the quantum ensemble, and how the corresponding volume profiles can be extracted. 
In Sec.\ \ref{sec:explo} we explore the phase diagram of the model numerically. 
In Sec.\ \ref{sec:tetdist} we study distributions of tetrahedra as a function of the coupling constants, which allows us
to understand how foliated (in the sense of regular CDT) typical configurations are.   
The results of our analysis of the volume profiles and their matching to a de Sitter universe are presented in 
Sec.\ \ref{sec:voldist}, and our conclusions in Sec.\ \ref{sec:conclusions}. --
Further details on the numerical implementation as well as a documentation of the relevant software 
can be found in \cite{thesis}.

\section{Review of Causal Dynamical Triangulations}
\label{cdt}
To set the stage for our subsequent generalization, we recall in the following some elements of Causal Dynamical 
Triangulations, mainly based on the original literature on CDT in 1+1 \cite{al}, 2+1 \cite{3dcdt} and 
3+1 dimensions \cite{cdt4d,desitter}.
For more extensive reviews and lecture notes on CDT we refer the interested reader 
to \cite{cdtreviews,physrep,cdtlectures}. 

\begin{figure}[t]
\centerline{\scalebox{0.55}{\includegraphics{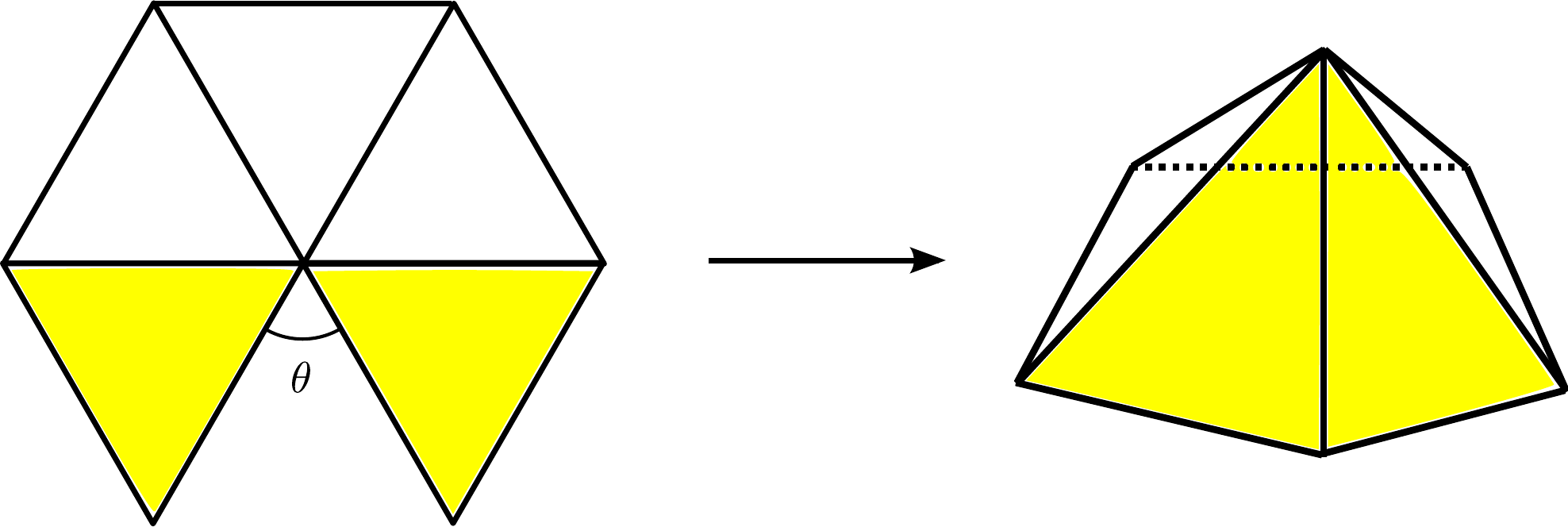}}}
\caption{A disc in the flat Euclidean plane is triangulated in terms of six equilateral triangles, 
one of which is subsequently removed (left). 
Regluing the remaining triangles one obtains a disc with positive curvature, characterized by a deficit angle
$\theta=\pi/3$ and independent of the embedding space (right).}
\label{fig:reggecurvature}
\end{figure}

The central object of interest in the CDT approach to quantum gravity is the gravitational path integral, 
which in the continuum can be written formally as
\begin{equation}
\label{eq:pigrav}
Z(G,\Lambda)=\hspace{-.6cm}\mathop{\int}_{\mathrm{geometries\, [g]}}\hspace{-.7cm}{\mathcal{D}[g]}\exp(iS_{\mathrm{EH}}[g]),
\end{equation}
where $S_{\mathrm{EH}}[g]$ is the Einstein-Hilbert action, written as functional of the metric $g$, $\mathcal{D}[g]$ is a measure 
on the space of geometries (the space of equivalence classes $[g]$ of metrics under the action of the diffeomorphism group), 
$G$ is Newton's constant and $\Lambda$ is the cosmological constant. 
To define the path integral properly it needs to be regularized, which in CDT is done by performing the ``sum over histories" 
(\ref{eq:pigrav}) over a set of piecewise flat, simplicial Lorentzian geometries -- in other words, triangulations -- effectively 
discretizing the curvature degrees of freedom of spacetime. 
The way in which triangulations encode curvature is illustrated best in two dimensions.
In the Euclidean plane, consider a flat disc consisting of six equilateral triangles which share a central vertex, and 
remove one of the triangles (Fig.\ \ref{fig:reggecurvature}, left). By identifying the opposite sides of the gap thus created, 
the piecewise flat disc acquires
nontrivial (positive Gaussian) curvature, whose magnitude is equal to the deficit angle $\pi/3$ at the vertex 
(Fig.\ \ref{fig:reggecurvature}, right). 
This also coincides with the rotation angle undergone by a two-dimensional vector parallel-transported around the vertex, and is 
therefore an {\it intrinsic} property of the two-dimensional disc. The principle of encoding curvature through deficit
angles (located at subsimplices of dimension $d-2$ in a $d$-dimensional triangulation) works in any dimension and
for any metric signature.

Let us review the difference between the (Euclidean) DT and the (Lorentzian) CDT path integral, again for 
simplicity in dimension two. Recall that the geometric properties of a flat triangle (or, in higher dimensions, a flat simplex)
are com\-pletely determined by the lengths of its edges.\footnote{To treat space-, time- and lightlike edges on the same footing, it is 
convenient to work with the {\it squared} edge lengths.}, including the {\it signature} of the flat metric in the building 
block's interior.  
In the Euclidean DT approach in two dimensions one works with a single type of building block, an
equilateral triangle of Euclidean signature, all of whose edges have some fixed spacelike length, 
the same length for all triangles in the
triangulation. As we have seen above, the number of such triangles around each interior vertex of the triangulation characterizes 
the local curvature at that vertex. In this way the formal integral over curved geometries in the continuum path integral (\ref{eq:pigrav}) 
is turned into a sum over triangulations, typically subject to some manifold conditions, which ensure that the triangulation
looks like a two-dimensional space everywhere.

\begin{figure}[t]
\centerline{\scalebox{0.7}{\includegraphics{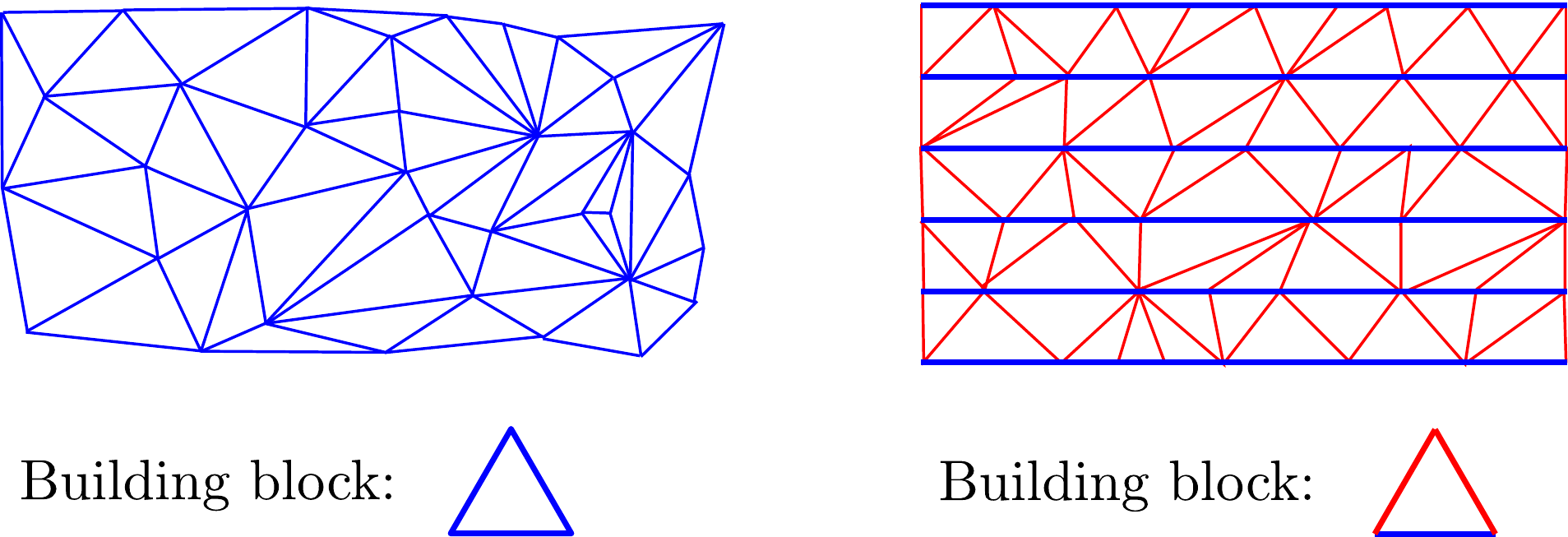}}}
\caption{Two-dimensional triangulations: from Euclidean DT, consisting of equilateral triangles with 
Euclidean signature (left), and from Lorentzian CDT, constructed from Minkowskian 
triangles with one spacelike and two timelike links (right). 
In both cases, curvature is present at all interior vertices whose coordination number (number of triangles meeting at the vertex) is
not six. Timelike links are shown in red, spacelike ones in blue.}
\label{fig:dt2d}
\end{figure}

By contrast, the flat triangles used in CDT quantum gravity have Lorentzian signature, something that cannot be achieved for 
equilateral triangles. The standard choice of an elementary CDT building block in 1+1 dimensions is one whose base
has squared length $\ell_s^2>0$ (and therefore is spacelike) and whose remaining two edges both have squared
length $\ell_t^2<0$ (and therefore are timelike). From the point of view of triangulating Lorentzian spacetimes,
one could in principle have chosen $\ell_t^2$ to be space- or lightlike, but
then CDT's prescription of a Wick rotation -- to be described below -- would no longer be applicable.  

Fig.\ \ref{fig:dt2d} illustrates the fundamental building blocks of two-dimensional DT (left) and CDT (right), and how they are
put together to obtain piecewise flat mani\-folds of Euclidean and Lorentzian signature. The two-dimensional graphs correctly
represent the neighbourhood relations of adjacent triangles, but not the actual length assignments, since it is impossible to 
flatten out a curved surface and preserve the edge lengths at the same time. Note that in the Lorentzian case the spacelike
edges form a sequence of one-dimensional simplicial submanifolds, which can be interpreted as hypermanifolds of
constant time $t=0,1,2, ...$, endowing each triangulation with a distinguished notion of (proper) time. This time
can be extended continuously to the interior of all triangles \cite{dl}. 

The ensemble of spacetimes forming the carrier
space of the CDT path integral are all triangulations which consist of a fixed number $t_{tot}$ of triangulated strips $\Delta t=1$,
where each strip is an arbitrary sequence of up- and down-triangles between times $t$ and $t+1$. The topology of space
(usually chosen to be $S^1$) is not allowed to change in time, that is, branchings into multiple $S^1$-universes are 
forbidden. It was shown in \cite{benedettihensonmatrix} that the global foliation of a 1+1 dimensional 
CDT spacetime into such strips can
be understood as consequence of a {\it local} regularity condition, namely, that precisely two spacelike edges be incident
on any vertex. Note that these geometries are causally well behaved and obey a piecewise linear analogue of
global hyperbolicity.  

\begin{figure}[t]
\centerline{\scalebox{0.6}{\includegraphics{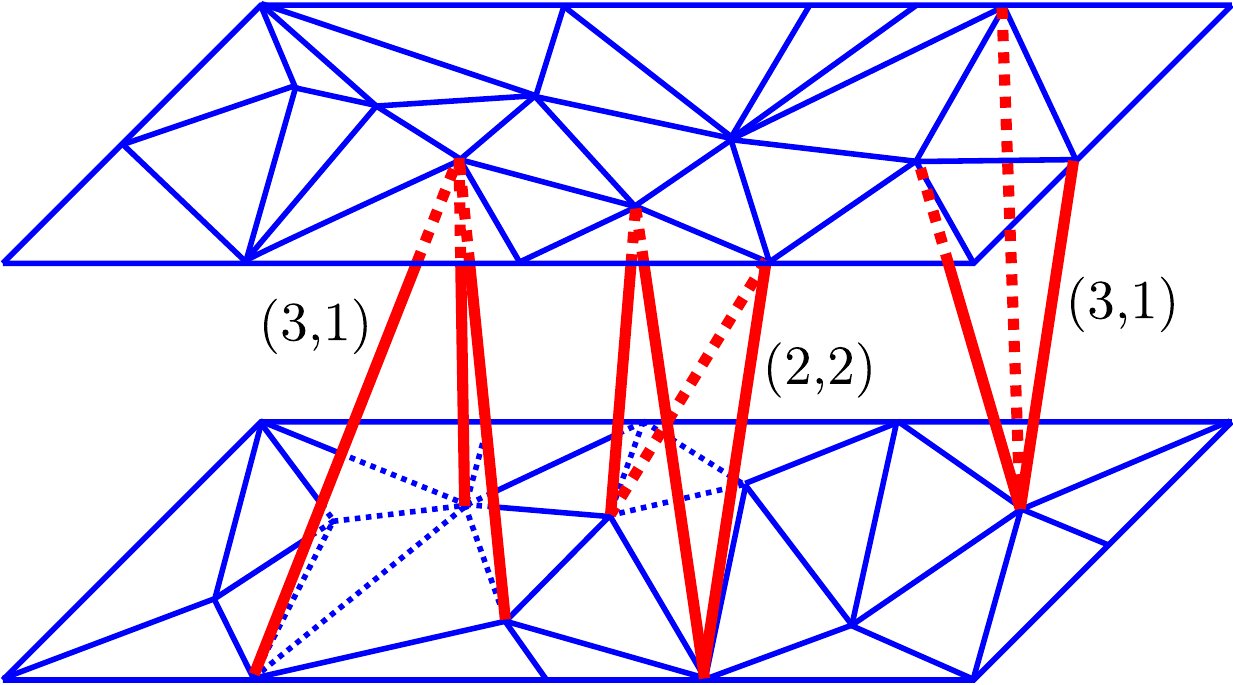}}}
\caption{Two adjacent spatial slices in a piece of foliated CDT triangulation in 2+1 dimensions. 
Tetrahedra of both types are shown.}
\label{fig:foliation}
\end{figure}

As already described in the introduction, the idea of creating individual path integral configurations with a 
well-behaved causal structure by imposing a preferred foliation on the underlying simplicial manifold  
is also realized in CDT in higher dimensions.
Fig.\ \ref{fig:foliation} shows part of a 2+1 dimensional CDT spacetime. Each leaf at integer-$t$ of the foliation forms a 
two-dimensional triangulation of the same fixed topology ${}^{(2)}\Sigma$, and consists of equilateral spacelike triangles 
with link length $\ell_s$. Adjacent triangulated spatial hypermanifolds are connected using Lorentzian tetrahedra to form 
a 2+1-dimensional simplicial manifold with spacetime topology $[0,1]\times {}^{(2)}\Sigma$, which again is a sequence of $t_{tot}$
triangulated ``slabs" $\Delta t=1$. Each such geometry comes with a time label $t$ which can be defined continuously
throughout the triangulation, and for integer values coincides with the discrete labeling of the simplicial leaves just 
described \cite{dl}. For reasons of simplicity, in the simulations time is often compactified. 
The topology then becomes $S^1\times {}^{(2)}\Sigma$.

The fundamental building blocks of three-dimensional CDT quantum gravity are flat Minkowskian tetrahedra, whose
geometric properties are determined by their edge lengths. As in 1+1 dimensions, one considers two different
link lengths, one spacelike with squared length $\ell_s^2$ and one timelike with $\ell_t^2=-\alpha \ell_s^2$.
Without loss of generality we have introduced here a positive constant $\alpha$, which quantifies the relative magnitude
of space- and timelike edge lengths and in what follows will be referred to as the \emph{asymmetry parameter}.

CDT path integral configurations are assembled from two different tetrahedron types, which can be distinguished  
by their orientation with respect to the preferred foliation. 
As can be seen from Fig.\ \ref{fig:foliation}, spacelike edges of a tetrahedron are always contained in a spatial 
submanifold of integer $t$, whereas timelike edges always connect different spatial slices. The (3,1)-tetrahedron
has three space\-like links forming a spacelike triangle, while the (2,2)-tetrahedron contains 
only two spacelike links. The notation $(i,j)$ indicates that $i$ vertices of the tetrahedron are located on one spatial slice 
and the remaining $j$ vertices on an adjacent one.

The kinematical set-up of CDT quantum gravity in 3+1 dimensions can be defined in a similar way. 
The leaves of the preferred foliation are three-dimensional 
Euclidean triangulations of fixed topology, and neighbouring slices are connected using Minkowskian four-simplices. 
Path integral configurations are simplicial manifolds assembled from 
two types of these building blocks, denoted by (4,1) and (3,2), depending on how their vertices are distributed 
among adjacent spatial slices. By labeling the foliation with increasing integers we again get a time variable for free, 
with every vertex being assigned a definite discrete time label.

Another ingredient that needs to be specified to make the model complete is an implementation 
on piecewise flat geometries of the Einstein-Hilbert
action in the path integral (\ref{eq:pigrav}), which in CDT quantum gravity is done following 
Regge's prescription 
\cite{regge}. One should point out that models of the type we are studying tend to be very robust with respect to
changes in the precise form of the action (which obviously is subject to discretization ambiguities) and of the configuration
space, in the sense that a wide range of different regularizations and kinematical ingredients will lead to the same 
continuum physics, if the latter can be defined meaningfully. An exception to this is of course the imposition of
causality constraints, which distinguishes CDT from DT quantum gravity and leads in all dimensions studied so far
to genuinely different continuum results. In two dimensions, this can be demonstrated exactly, for example, by 
comparing specific observables and critical exponents, since the CDT model 
can be solved analytically \cite{al}. In higher dimensions, information about the behaviour of observables comes
primarily from numerical simulations: three-dimensional CDT quantum gravity has only been solved partially and 
for restricted classes of triangulations \cite{blz}, while in four dimensions analytical methods are mostly unavailable 
and one must resort to Monte Carlo simulations to extract physical results.

\begin{figure}[t]
\centerline{\scalebox{0.6}{\includegraphics{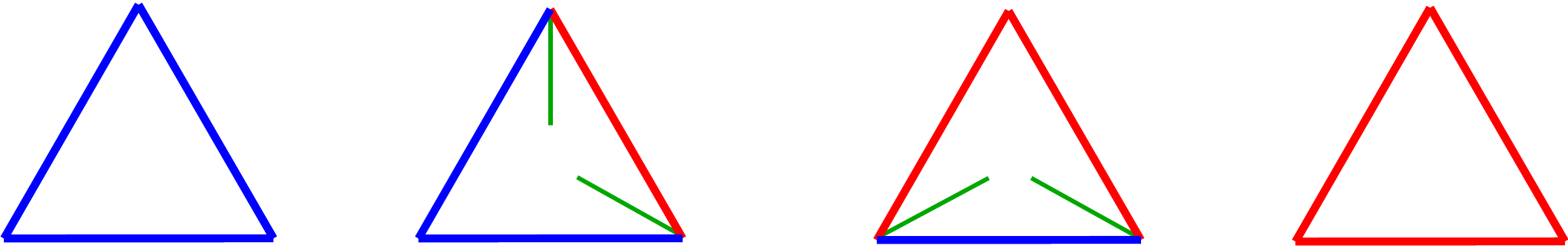}}}
\caption{The four triangle types which can be constructed from using just two link lengths, spacelike (blue) and timelike (red). 
Only the two Minkowskian triangles (of types $sst$ and $tts$) at the centre have the correct signature for 
triangulating 2d Lorentzian spacetimes. -- The green lines inside the Lorentzian triangles indicate light rays 
through the corner vertices.}
\label{fig:fourtriangles}
\end{figure}
 
In order to analyze the dynamics of CDT quantum gravity using such simulations, a Wick rotation must be performed
to convert the complex path integral amplitudes to real Boltzmann weights. This can be achieved by performing an 
analytic continuation of the asymmetry parameter, by rotating it in the lower half of the complex plane such that 
$\alpha$ is mapped to $-\alpha$ \cite{3d4d}. As a consequence, the gravitational path integral  
becomes a statistical partition function of the form
\begin{equation}
\label{partition}
Z=\sum_{T\in \cal{C}}\frac{1}{C(T)}\exp(-S_{\mathrm{Regge}}^{\mathrm{eucl}}(T)),
\end{equation}
where $\cal{C}$ is the space of all causal, Lorentzian triangulations $T$, $S_{\mathrm{Regge}}^{\mathrm{eucl}}$ 
the Euclideanized 
Regge action and $1/C(T)$ the discrete analog of the path integral measure, with $C(T)$ denoting the order of the 
automorphism group of $T$. In Sec.\ \ref{sec:actions} below we will derive and discuss the explicit 
functional form of the Regge implementation of the three-dimensional Einstein-Hilbert action in terms of the 
triangulation data and the coupling constants of the nonfoliated CDT model. 

\section{Relaxing the foliation: 1+1 dimensions}
\label{sec:ncdt_2d}

As a warm-up for the three-dimensional case, we will in this section illustrate our general strategy for relaxing the distinguished
foliation by discussing the situation in 1+1 dimensions. The key idea is to add new elementary Minkowskian building blocks, while
sticking to two types of links, one spacelike and one timelike, where we will continue to use the notation $\ell_s^2$ and
$\ell_t^2\! =\!-\alpha \ell_s^2$ for their squared lengths. Fig.\ \ref{fig:fourtriangles} shows the four types of triangles which can be 
built from these two link types. By calculating the metric inside the triangles one finds that there are exactly two which have
Lorentzian signature $(-+)$, the two at the centre of the figure. We will call them the $sst$- and the $tts$-triangle respectively, in
reference to the spacelike ($s$) and timelike ($t$) edges they contain. Note that in standard CDT in two dimensions only the
$tts$-triangle is used.

\begin{figure}[t]
\centerline{\scalebox{0.65}{\includegraphics{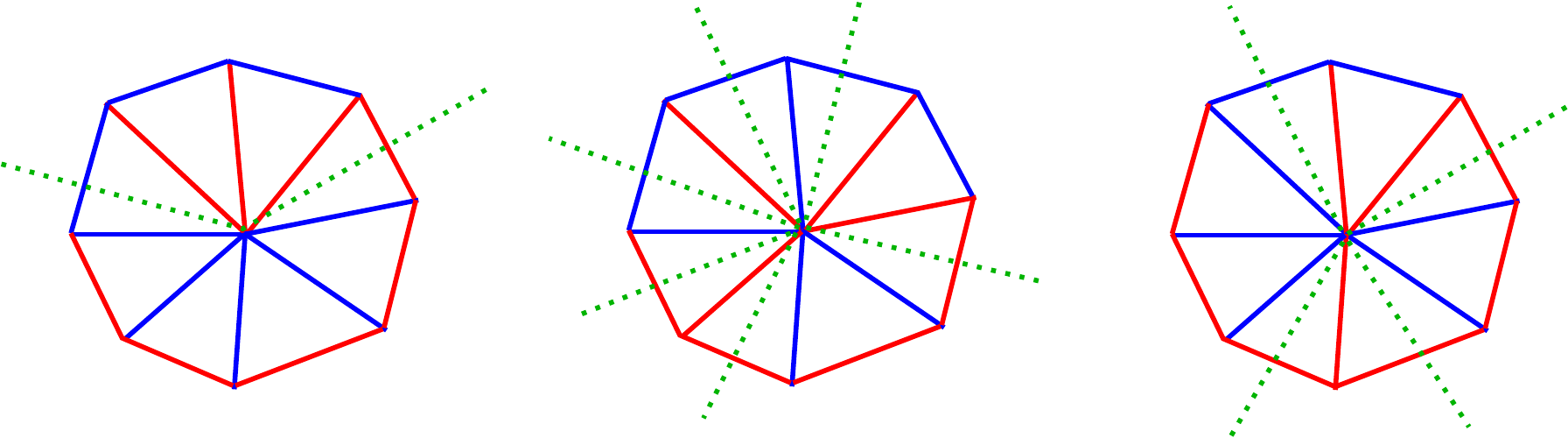}}}
\caption{Three examples of vertex neighbourhoods in 1+1 dimensions. Causality is violated at the central vertex whenever 
the number of light cones is not two (left and centre). At a causally well-behaved vertex, one crosses exactly two light cones
(equivalently, four lightlike directions) when going around the vertex (right).}
\label{fig:causality2d}
\end{figure}

The use of Lorentzian building blocks is not a sufficient condition for the triangulation to have a well-defined causal
structure locally, we also need to check that we obtain well-defined light cones in points where triangles  
are glued together. If the gluing happens according to the standard rule of only identifying links of the same type
(spacelike with
spacelike, timelike with timelike), the only local causality violations\footnote{Whenever we talk about geometries being
{\it causal}, what we have in mind is that they possess a well-behaved {\it causal structure}. This should
not be confused with the notion of causality in standard (quantum) field theory, which refers to the behaviour of matter 
fields on a given background that typically already comes with a fixed causal structure.} can occur 
at the vertices of the triangulation.
Counting past and future light cones separately, the point is that one may obtain more or fewer than the required 
two light cones at a vertex, as illustrated by the local neighbourhoods depicted in Fig.\ \ref{fig:causality2d}. 
Local causality implies crossing exactly two light cones when going around a vertex or, equivalently, crossing
exactly four lightlike lines emanating radially from the vertex.

\begin{figure}[t]
\centerline{\scalebox{0.55}{\includegraphics{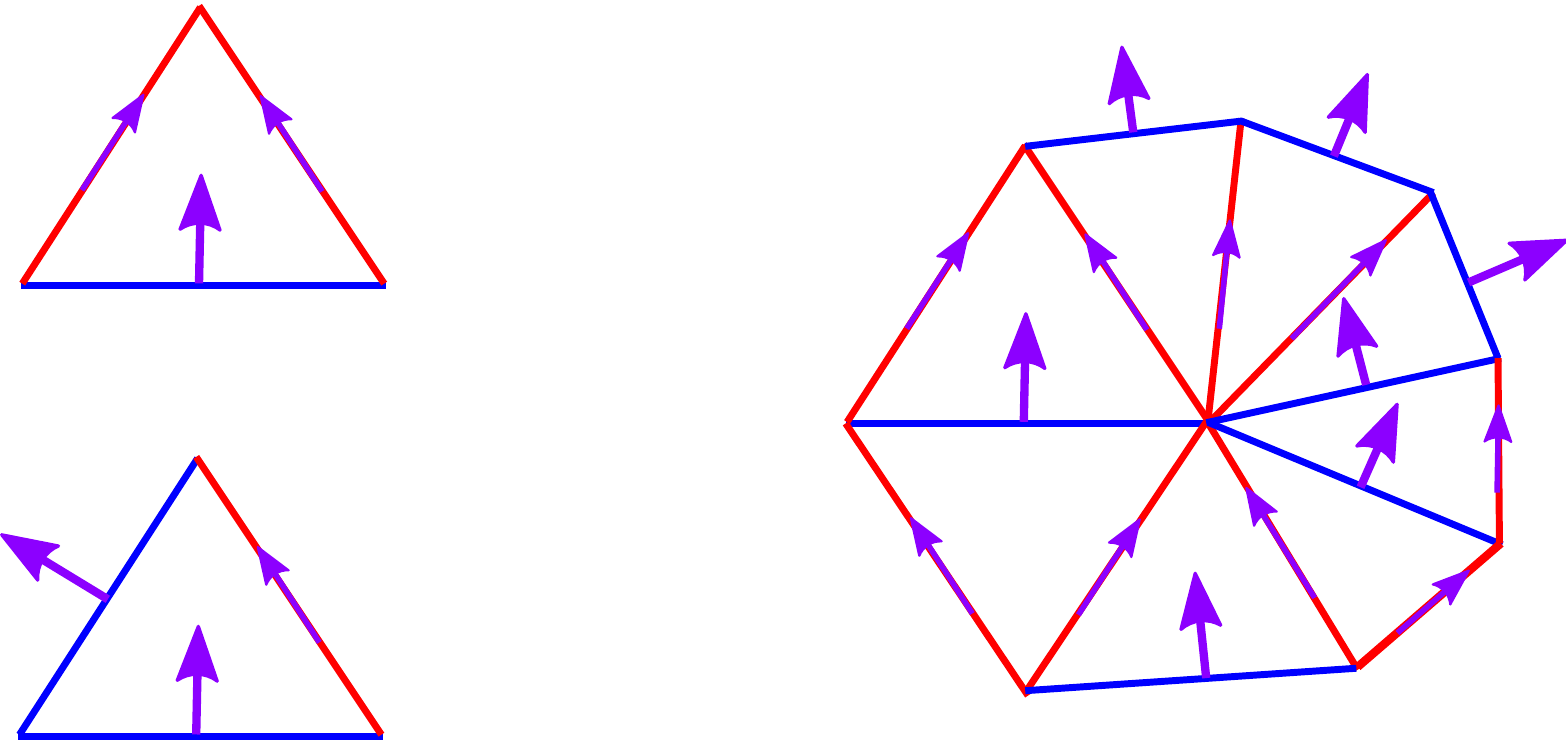}}}
\caption{Two Lorentzian triangles with consistent assignments of future-pointing arrows to its edges, 
as explained in the text (left). 
The time orientation of a given triangle determines the time orientation of its direct neighbours (right).}
\label{fig:flowoftime2d}
\end{figure}

As a next step, we will consider the global causal structure of individual triangulated manifolds 
satisfying the local vertex causality condition everywhere. This global structure will in general depend on the chosen topology.
For simplicity, we will restrict ourselves to the cases $[0,1]\times S^1$ (space is a compact circle) 
and $[0,1]\times [0,1]$ (space
is a closed interval), where the initial and final boundary are assumed to be spacelike, and any other boundaries (in the second
case) timelike. We will call such a spacetime globally causal if it does not contain any closed timelike curves. 
A ``timelike curve" for our purposes will be a sequence of oriented, timelike links in a triangulation. Since the interior of
every Minkowskian triangle has a well-defined light cone structure, a choice of orientation (i.e. a choice of which one
is the past and which one the future light cone) induces an orientation on its timelike edges, which can be captured
by drawing a future-pointing arrow onto the edge. Conversely, these arrow assignments fix the orientation of the
triangle uniquely. To follow the ``flow of time" in a triangulation it is convenient to also associate a future-pointing 
arrow with each spacelike edge, which is drawn perpendicular to the edge, see Fig.\ \ref{fig:flowoftime2d}. 

Choosing a consistent orientation for all building blocks in standard CDT in this way is completely straightforward:
each triangle sits in a strip between discrete times $t$ and $t+1$, which fixes its orientation uniquely. Independent
of the spatial boundary conditions, there are no closed timelike curves (unless we impose periodic boundary
conditions {\it in time}, which trivially makes any timelike curve closed, a situation we are not considering here). 

By contrast, the situation in nonfoliated CDT
is slightly more involved. Given a time-oriented triangle, the orientation of a neighbouring triangle that shares an
edge with the first one is uniquely determined by consistency. It is easy to see that when vertex causality is violated
(like in the example of Fig.\ \ref{fig:causality2d}, left), inductively assigning orientations in this way will fail -- i.e. lead to
contradictions -- even for a local vertex neighbourhood. If vertex causality is satisfied, one can show that for noncompact
spatial topology there are no closed timelike curves \cite{hoekzema}. For compact spatial slices, where the spacetime
topology is that of a cylinder, one can construct explicit geometries which exhibit noncontractible, closed timelike curves. 
Of course, we do not know a priori whether the presence of closed timelike curves in individual path integral configurations
has any influence on the continuum limit of the model, and perhaps leads to undesirable continuum properties.
It appears that in the context of our present investigation this issue is largely circumvented. Although the causality
conditions we impose are of a local nature, and may admit the presence of closed timelike curves, it turns out that
the geometries dominating the sum over histories dynamically retain a weak degree of foliation 
(see Sec.\ \ref{sec:tetdist} below), which suggests that such curves are certainly not abundant. We have not seen
closed timelike curves in random samples, but have not systematically tested for their presence either.
 
\begin{figure}[t]
\centerline{\scalebox{1.1}{\includegraphics{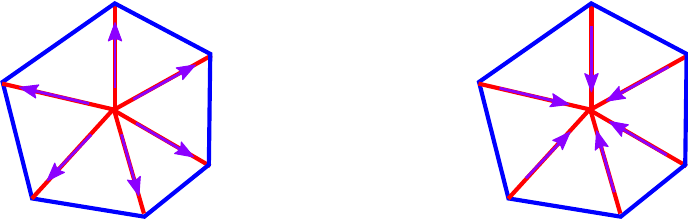}}}
\caption{A source and a sink of time in 1+1 dimensions. In both cases, vertex causality is violated at
the central vertex.}
\label{fig:sourcesink2d}
\end{figure}

Anticipating the choice of boundary conditions we will make in 2+1 dimensions, we
may relax the local causality constraint slightly by allowing for the presence of an isolated
``source" and ``sink" of time. By this we mean a vertex where only timelike links meet, all of them either
time-oriented away from the
vertex (source) or toward it (sink), as illustrated by Fig.\ \ref{fig:sourcesink2d}. For compact spatial boundary
conditions, choosing a source and a sink as initial and final (degenerate spatial) boundaries 
will convert the cylinder into a
spherical $S^2$-spacetime topology. A similar choice of boundary conditions in 2+1 dimensions will lead to a
$S^3$-spacetime topology, with the source and sink forming the south and north pole of the sphere, as we will
see later.

\begin{figure}[b]
\centerline{\scalebox{0.4}{\includegraphics{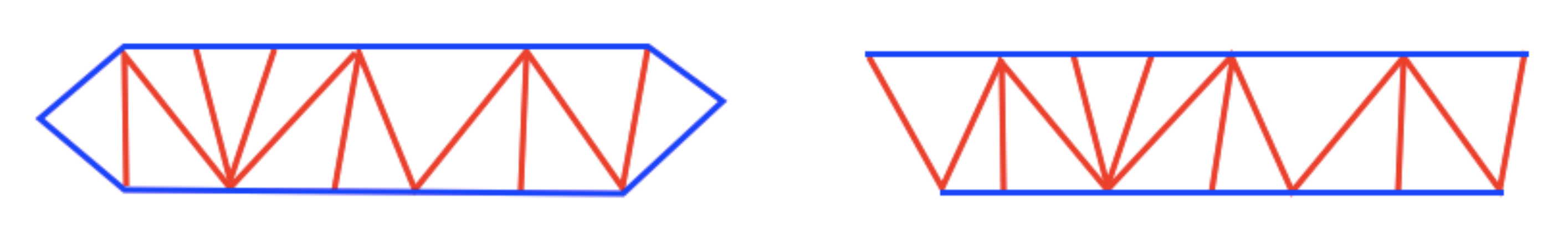}}}
\caption{A bubble (left) and a strip configuration (right) in 1+1 dimensions.}
\label{fig:bubblestrip}
\end{figure}

There is a particular substructure of the triangulations, called a {\it bubble} \cite{hoekzema}, 
which involves the newly 
added building blocks and is helpful in analyzing the geometry of nonfoliated CDT. 
In 1+1 dimensions it is simply a pair of $sst$-triangles with a chain of $tts$-triangles in between. 
This is the general structure of a two-dimensional connected region bounded by a closed loop of spacelike links, 
and whose interior contains only timelike links, as shown in Fig.\ \ref{fig:bubblestrip}, left (we are assuming that vertex causality is 
satisfied everywhere). This should be contrasted with the structure of a strip, which likewise denotes a 
two-dimensional piece of triangulations bounded by spacelike links and without spacelike links in its interior, but
whose boundary is disconnected (Fig.\ \ref{fig:bubblestrip}, right). CDT quantum gravity in 1+1 dimensions has only strips, whereas 
the version without distinguished foliation has both strips and bubbles. Analogous structures will play a role
in our analysis in 2+1 dimensions too, where also the interior structure of a bubble can become more complicated.

\section{Relaxing the foliation: Kinematics in 2+1 dimensions}
\subsubsection*{Local causality conditions}
\label{sec:nfct21}

\begin{figure}[t]
\centerline{\scalebox{0.55}{\includegraphics{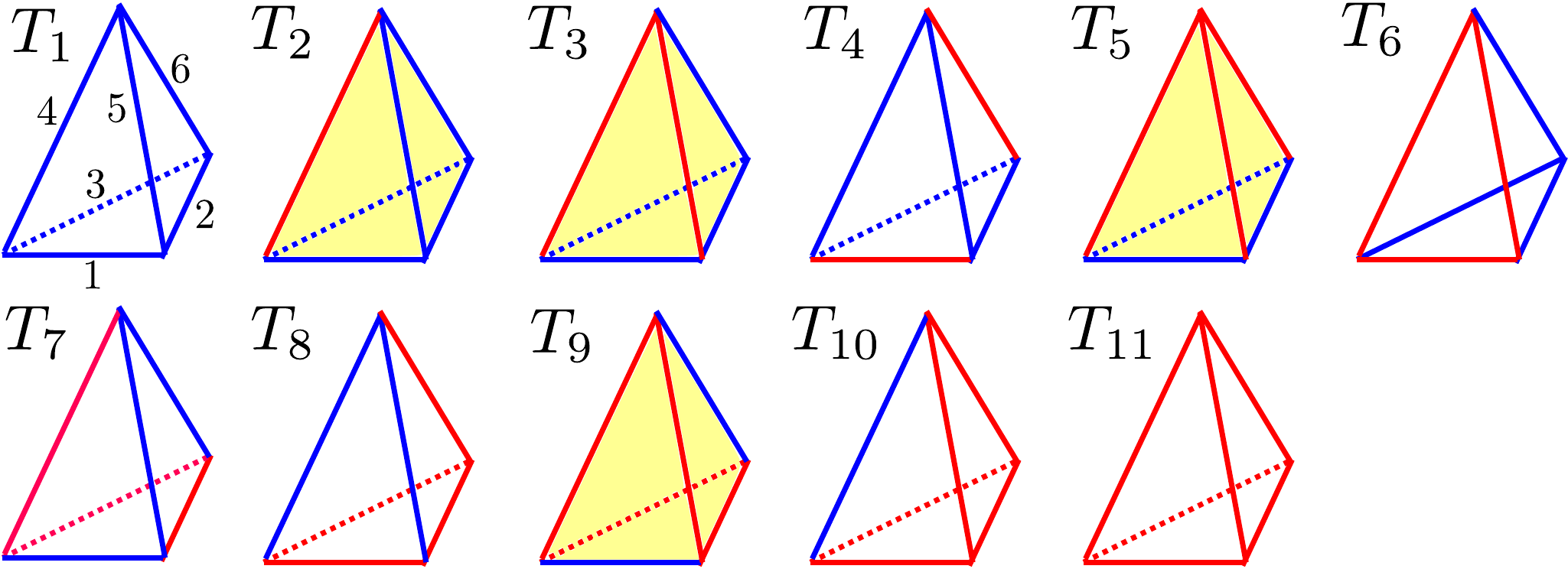}}}
\caption{All tetrahedra in 2+1 dimensions that can be constructed from time- and spacelike links of fixed squared length, 
allowing for any signature. The tetrahedra highlighted in yellow have the correct Lorentzian signature; for 
the $T_7$-tetrahedron this only holds for $\alpha<1$. The link labeling shown for the first tetrahedron will be used
for the other tetrahedra too.}
\label{fig:tetrahedra}
\end{figure}

Following an analogous procedure in 2+1 dimensions to arrive at a model without distinguished simplicial hypermanifolds,
we first must determine which flat tetrahedra -- again only built from two types of edge lengths -- give rise to
Minkowskian building blocks of the correct Lorentzian signature $(-++)$. Fig.\ \ref{fig:tetrahedra} shows all types of 
tetrahedra which can be constructed using space- and timelike links with fixed squared lengths $\ell_s^2$ and
${\ell_t^2\! =\! -\alpha \ell_s^2}$ respectively. In the rest of this document we will set $\ell_s\! =\! 1$. 
By calculating the metric in the interior of each tetrahedron type, one finds that only $T_2$, $T_3$, $T_5$ and $T_9$ have
the required signature for all values $\alpha\! >\! 0$ of the asymmetry parameter, and type $T_7$ only for $0\! <\! \alpha\! <\! 1$. 
Note that standard CDT quantum gravity only uses the tetrahedra $T_5$ and $T_9$, in Sec.\ \ref{cdt} referred to as (3,1)- 
and (2,2)-tetrahedra respectively.
In the present work we will for reasons of simplicity
investigate the version of the model where causal spacetimes are assembled from the tetrahedral
types $T_2$, $T_3$, $T_5$ and $T_9$ (without $T_7$). As we will see, this already serves our purpose of breaking up 
the fixed foliated structure.   

\begin{figure}[t]
\centerline{\scalebox{0.7}{\includegraphics{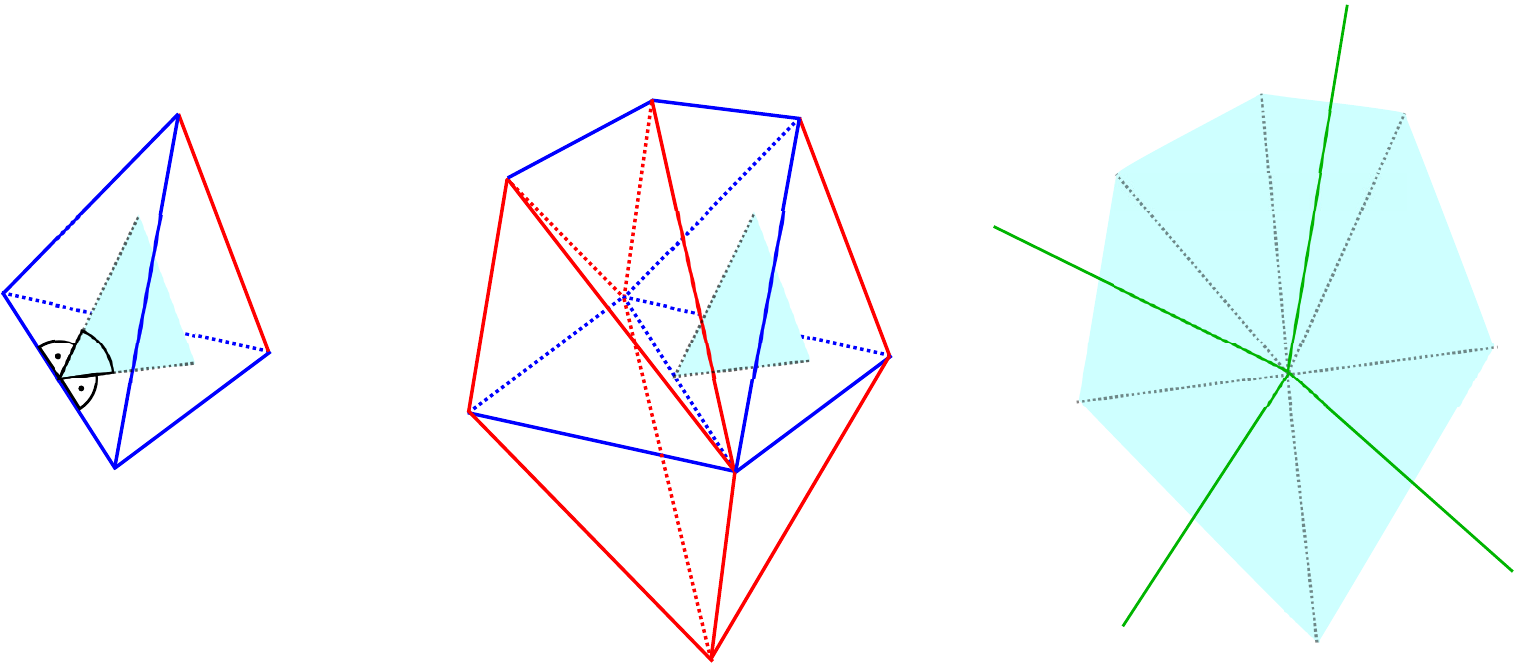}}}
\caption{Dihedral angle at the link with label ``2" of a $T_2$-tetrahedron (left). Star of a link, all of whose
tetrahedra contribute dihedral angles at the link (centre). Two-dimensional cut perpendicular to the link. In the case
depicted, the space is Lorentzian, with green lines representing light rays originating at the centre (right).}
\label{fig:causality3d}
\end{figure}

In 2+1 dimensions violations of local causality -- which should therefore be forbidden by the gluing rules --
can in principle occur at the links and the vertices of a triangulation. To check whether the light cone structure at
a given link is well-behaved, it is sufficient to consider the geometry of a two-dimensional piecewise flat
surface orthogonal to the link at its midpoint. This geometry is completely characterized by 
the set of tetrahedra sharing the link, the so-called star of the link (Fig.\ \ref{fig:causality3d}, centre).
Each tetrahedron in the star contributes a dihedral angle, defined by the intersection of the tetrahedron with
the plane perpendicular to the link (Fig.\ \ref{fig:causality3d}, left).    
The plane segments spanned by all the dihedral angles associated to the given link form a new plane\footnote{This is
a slight misnomer; in general, this ``plane" will not be flat, because the vertex at the centre will carry a nonvanishing
deficit angle.}, as shown in Fig.\ \ref{fig:causality3d} (right).

We can distinguish between two cases. If the link at the centre of the star is timelike, the metric of the plane 
has Euclidean signature, all dihedral angles are Euclidean, and there are no further causality conditions to be satisfied.
If on the other hand the link is spacelike, the orthogonal plane is Lorentzian, and so are the dihedral angles. Like in the
1+1 dimensional case discussed in the previous section, we must then require that there is exactly one
pair of light cones at the central vertex, and that we encounter exactly four lightlike directions when
circling around it. We say that the triangulation satisfies \emph{link causality} if this condition is satisfied for 
every spacelike link.

\begin{figure}[t]
\centerline{\scalebox{0.65}{\includegraphics{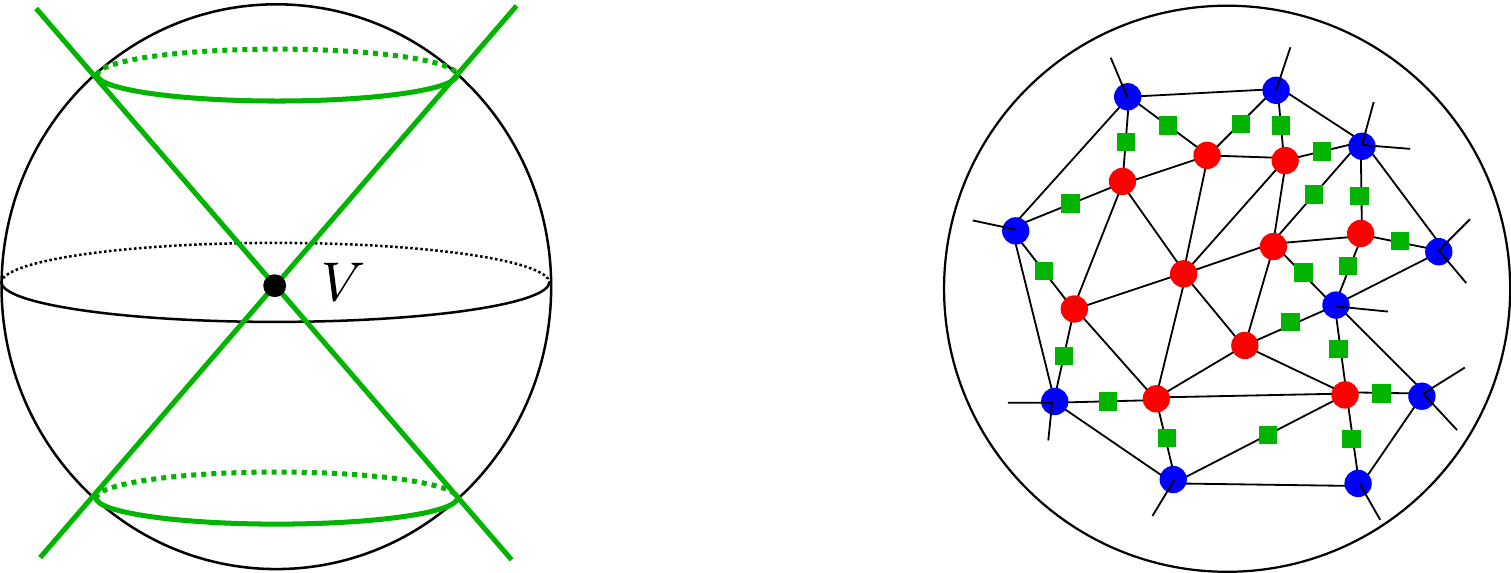}}}
\caption{In a triangulation obeying vertex causality, intersecting the boundary of a spherical vertex neighbourhood 
with the light cones at the vertex $V$ results in two disconnected circles (left). 
Part of the surface triangulation of a unit ball around $V$, 
showing timelike radial links as red, spacelike ones as blue and light cone crossings as green dots (right).}
\label{fig:vertexcausality}
\end{figure}

Link causality guarantees that light cones everywhere in the triangulation are regular, except possibly at vertices. 
Intersecting the light cone(s) at a vertex $V$ with the surface of a unit ball around $V$, we obtain 
two disconnected circles if and only if local causality holds at $V$, see Fig.\ \ref{fig:vertexcausality} (left). 
In terms of the triangulated surface $\cal S$ of the unit neighbourhood around $V$, vertex causality can be characterized as
follows. Mark the end of a timelike link between $V$ and $\cal S$ by a red dot and that of a spacelike one by a blue dot.
In addition, whenever the light cone through $V$ crosses a link on $\cal S$, mark the link with a green dot. 
Recalling the situation depicted in Fig.\ \ref{fig:fourtriangles}, it is clear that a green dot will always occur on a
surface link which connects a red and a blue dot. If we cut all links that are marked with a green dot, the surface 
triangulation will break up into a number of connected components.
If two of the components thus obtained contain red vertices and one component contains blue vertices, we say that 
\emph{vertex causality} holds at $V$. If this is true for all vertices, we say that the triangulation satisfies 
vertex causality.
We have not found any Monte Carlo moves which destroy vertex causality but maintain link causality. 
Also, we have not been able to explicitly construct a triangulation that satisfies link causality and violates vertex 
causality, but we do not currently have a proof that link causality implies vertex causality.

In order to compute the explicit action for the generalized CDT model of 2+1 dimensional quantum gravity, we will need the 
values of all dihedral angles. As usual, we will use Sorkin's complex angle prescription \cite{sorkin} for the latter, which 
conveniently keeps track of both Euclidean and Lorentzian angles. The analytic expressions for the cosines and sines of the 
dihedral angles are listed in Table\ \ref{tab:tetraangles}, from which the angles can be computed uniquely. Closer 
inspection of the geometry of the tetrahedra reveals that a dihedral angle contains a light cone crossing whenever
the two triangles bounding the angle are a pair of a spacelike (Euclidean) and a Lorentzian triangle. 
Local link causality therefore implies that the triangle type changes exactly four times between spacelike and non-spacelike 
when we circle around a spacelike link once. The number of light cone crossings in the case of a Lorentzian angle is
also contained in the table.

\begin{table}[t]
\begin{center}
\renewcommand{\arraystretch}{1.6}
\begin{tabular}{|c|c|c|c|c|}
\hline
tetrahedron & links & $\cos(\Theta)$ & $\sin(\Theta)$ & light cone crossings\\
\hline
\hline
$T_2$ & $1,3,5,6$ & $\frac{i}{\sqrt{3}}\sqrt{\frac{\alpha}{4+\alpha}}$ & $\frac{2}{\sqrt{3}}\frac{\sqrt{3+\alpha}}{\sqrt{4+\alpha}}$ &  $1$ \\
& $2$ & $1+\frac{2\alpha}{3}$ & $\frac{2i\sqrt{\alpha(3+\alpha)}}{3}$ & $0$ \\
& $4$ & $\frac{2+\alpha}{4+\alpha}$ & $\frac{2\sqrt{3+\alpha}}{4+\alpha}$ & $-$ \\
\hline
$T_3$ & $1$ & $\frac{i(1+2\alpha)}{\sqrt{3+12\alpha}}$ & $2\sqrt{\frac{1+\alpha(4+\alpha)}{3+12\alpha}}$ & $1$ \\
& $2,3$ & $\frac{2+\alpha}{\sqrt{3}\sqrt{-\alpha(4+\alpha)}}$ & $\frac{2}{\sqrt{3}}\sqrt{\frac{1+\alpha(4+\alpha)}{\alpha(4+\alpha)}}$ & $1$ \\
& $4,5$ & $\frac{1}{\sqrt{17+\frac{4}{\alpha}+4\alpha}}$ & $\frac{2}{\sqrt{4+\frac{\alpha}{1+\alpha(4+\alpha)}}}$ & $-$ \\
& $6$ & $\frac{2+\alpha(4+\alpha)}{\alpha(4+\alpha)}$ & $-\frac{2i\sqrt{1+\alpha(4+\alpha)}}{\alpha(4+\alpha)}$ & $0$ \\
\hline
$T_5$ & $1,2,3$ & $-\frac{i}{\sqrt{3}\sqrt{1+4\alpha}}$ & $\frac{2}{\sqrt{3}}\frac{\sqrt{1+3\alpha}}{\sqrt{1+4\alpha}}$ & $1$ \\
& $4,5,6$ & $\frac{1+2\alpha}{1+4\alpha}$ & $\frac{2\sqrt{\alpha(1+3\alpha)}}{1+4\alpha}$ & $-$ \\
\hline
$T_9$ & $1,6$ & $\frac{3+4\alpha}{1+4\alpha}$ & $-\frac{2i\sqrt{2+4\alpha}}{1+4\alpha}$ & $0$ \\
& $2,3,4,5$ & $-\frac{1}{1+4\alpha}$ & $\frac{2\sqrt{2}\sqrt{\alpha(1+2\alpha)}}{1+4\alpha}$ & $-$ \\
\hline
\end{tabular}
\end{center}
\caption{Dihedral angles $\Theta$ for all tetrahedra types, given in terms of their trigonometric functions. 
The link numbers refer to the numbering given
in Fig.\ \ref{fig:tetrahedra}. For Lorentzian angles, also the number of light cone crossings is given.}
\label{tab:tetraangles}
\end{table}

\begin{figure}[b]
\centerline{\scalebox{0.70}{\includegraphics{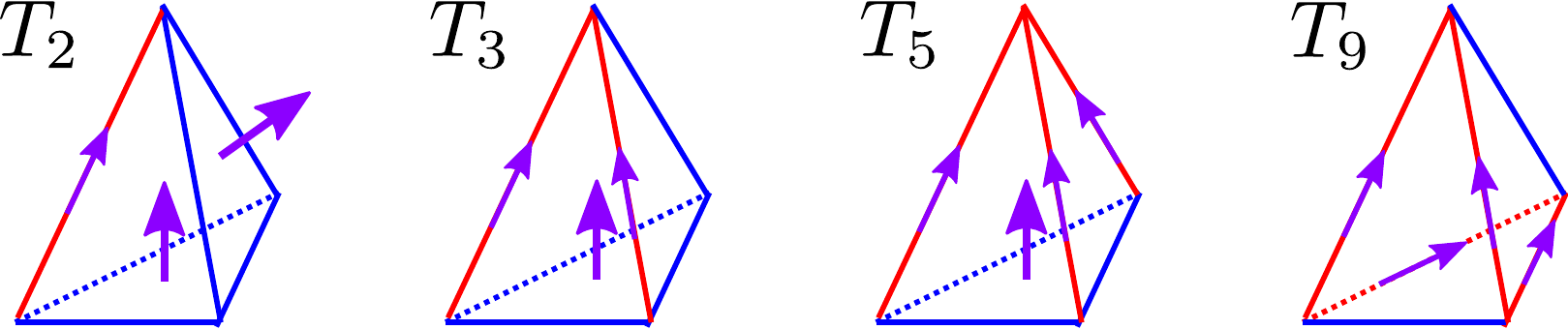}}}
\caption{The four fundamental tetrahedral building blocks, each equipped with one (out of two
possible) time orientations.}
\label{fig:flowoftime3d}
\end{figure}

Just like in 1+1 dimensions, choosing a time-orientation for a tetrahedron induces an orientation on its timelike links,
as well as on the normal to any of its spacelike triangles. It is operationally convenient to keep track of these data
in terms of future-oriented arrow assignments, as illustrated by Fig.\ \ref{fig:flowoftime3d}. 
Again, the local causality conditions do not guarantee that the time orientation can be extended to the full triangulation. 
In addition to these conditions, we will therefore require (and enforce by way of our computer algorithm) 
that the complete triangulation can be time-oriented consistently. 

\subsubsection*{Simplicial substructures}

In trying to understand the local geometry of nonfoliated CDT configurations and how it is affected by the Monte Carlo
moves defined in Sec.\ \ref{sec:numsetup} below, it is useful to isolate specific local substructures built from the 
fundamental tetrahedra of Fig.\ \ref{fig:flowoftime3d}.
To start with, note that only tetrahedra of type $T_2$ and $T_3$ contain triangles with exactly two spacelike edges. 
Furthermore, both tetrahedra have exactly two such triangles. If we glue two of them together along such a triangle, 
the resulting simplicial complex again has two such triangles on its boundary. 
Iterating this gluing procedure we end up with a chain of tetrahedra of type $T_2$ and $T_3$. 
We conclude that in a triangulation without boundary, the set of all tetrahedra of type $T_2$ and $T_3$ 
necessarily organizes itself into a collection of closed rings. In a triangulation with boundary also open chains 
are possible.

\begin{figure}[t]
\centerline{\scalebox{0.6}{\includegraphics{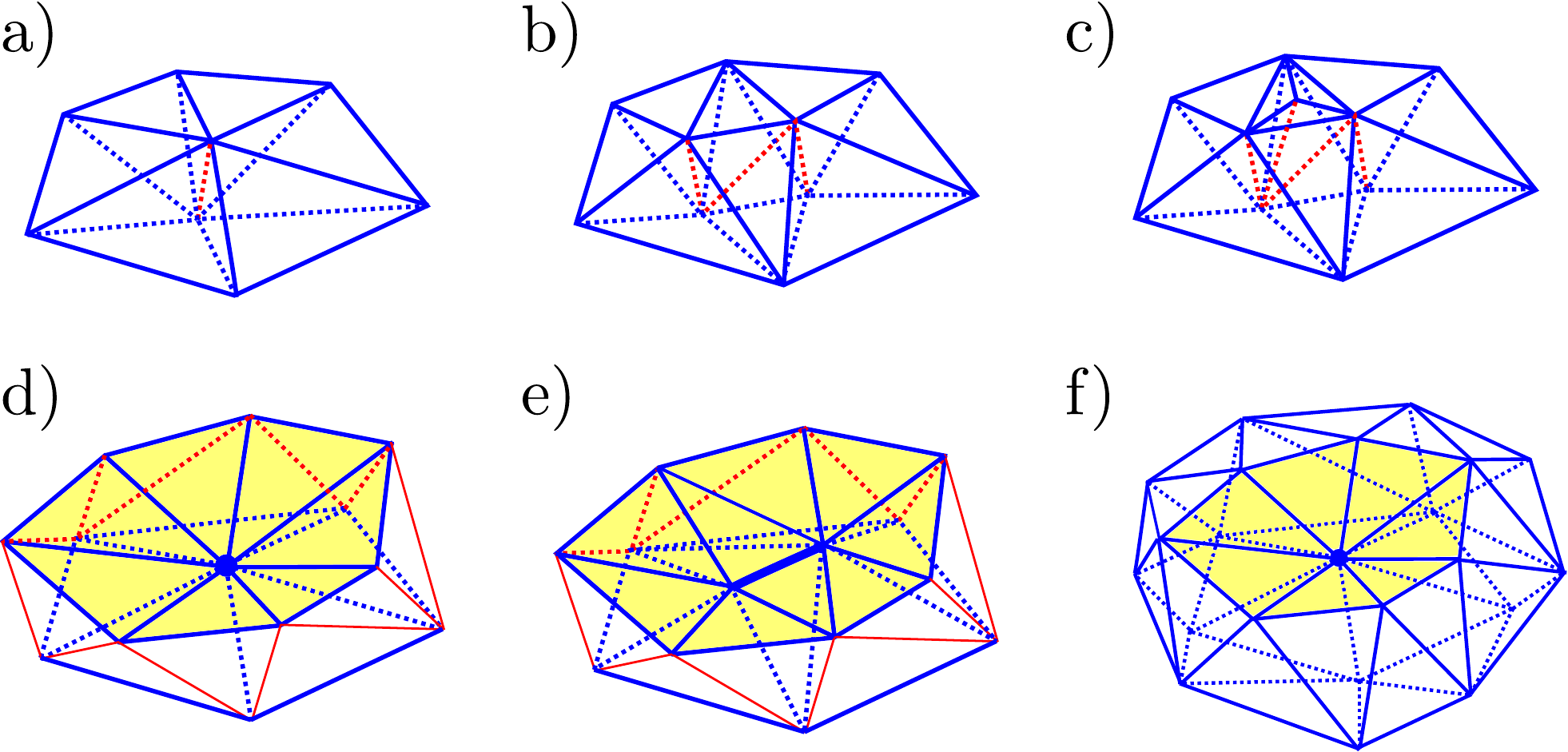}}}
\caption{(a) A ring of $T_2$-tetrahedra, forming a simple bubble. (b) By inserting tetrahedra of type $T_3$, 
we can form more complicated bubbles. (c) More general bubbles also contain other tetrahedra types. 
(d) A ring of $T_3$-tetrahedra gives rise to a pinching, where two spatial discs meet in a single vertex. 
(e) By adding two $T_2$-tetrahedra, the pinching is extended to a link. 
(f) Bubbles and pinchings can occur in combination (timelike links are omitted here).}
\label{fig:rings}
\end{figure}

Using these simplicial substructures, we can construct three-dimensional analogues of the ``bubbles" 
of Sec.\ \ref{sec:ncdt_2d} above, by which we will mean connected pieces of triangulation enclosed by a
surface made of only spacelike triangles, with no such triangles in its interior. 
If a ring only contains tetrahedra of type $T_2$, we get a simple bubble, 
consisting of two spatial discs with identical structure and a timelike link in its interior (Fig.\ \ref{fig:rings}a.)
By inserting $T_3$-tetrahedra into a ring of $T_2$-tetrahedra we can form more complicated bubbles, as illustrated
by Fig.\ \ref{fig:rings}b. More general bubbles consist of an outer ring of $T_2$- and $T_3$-tetrahedra, enclosing one
or more tetrahedra of the other two types, like the one shown in Fig.\ \ref{fig:rings}c.

We can also consider a ring of $T_3$-tetrahedra, as depicted in Fig.\ \ref{fig:rings}d. The spacelike triangles marked in 
yellow form a spatial disc, with a similar spacelike disc just below. Both discs meet in a single vertex, their
respective centres, which we will refer to as a {\it pinching} at that vertex.
This situation can be generalized by inserting tetrahedra of type $T_2$ into the $T_3$-ring, as shown in 
Fig.\ \ref{fig:rings}e. The effect is that the two spatial discs now intersect in a link rather than a vertex. 
Bubbles and pinchings can occur in combination to create even more complicated structures, 
an example of which is shown in Fig.\ \ref{fig:rings}f. A feature of bubbles which we have not yet mentioned is
that they can self-overlap, in the sense that the spherical (or possibly higher-genus) surface bounding a bubble 
may touch itself along some subset of the surface triangulation. As explained below, 
we will exclude one kind of self-overlapping bubbles from our simulations,
namely, those that wrap nontrivially around the spatial two-sphere.

\subsubsection*{Kinematical constraints}
\label{sec:constraints}

The simplest information one can extract from a triangulation is the number of its subsimplices of a particular type. 
We will use the following counting variables for the four fundamental tetrahedra and
the lower-dimensional subsimplex types, as well as their sums in each dimension:
\begin{align}
\label{countvar}
N_0&=\textrm{number of vertices}\nonumber \\
N_1^s&=\textrm{number of spacelike links}\nonumber \\
N_1^t&=\textrm{number of timelike links}\nonumber \\
N_1&:=N_1^s+N_1^t \nonumber\\
N_2^{sss}&=\textrm{number of triangles with three spacelike links}\nonumber \\
N_2^{sst}&=\textrm{number of triangles with two spacelike links}\nonumber \\
N_2^{tts}&=\textrm{number of triangles with one spacelike link} \\
N_2&:=N_2^{sss}+N_2^{sst}+N_2^{tts} \nonumber \\
N_3^{T_2}&=\textrm{number of tetrahedra of type}\ T_2 \nonumber\\
N_3^{T_3}&=\textrm{number of tetrahedra of type}\ T_3 \nonumber \\
N_3^{T_5}&=\textrm{number of tetrahedra of type}\ T_5 \nonumber \\
N_3^{T_9}&=\textrm{number of tetrahedra of type}\ T_9 \nonumber\\
N_3&:=N_3^{T_2}+N_3^{T_3}+N_3^{T_5}+N_3^{T_9}.\nonumber
\end{align}

There exist linear identities among these numbers, which for CDT have been described in \cite{3d4d}. 
Here we will repeat the analysis for the extended ensemble, including the new tetrahedral building blocks. 
The first identity,
\begin{equation}
\label{id1}
N_0-N_1+N_2-N_3=\chi,
\end{equation}
involves the Euler characteristic $\chi$ of the simplicial spacetime manifold. Since every tetrahedron contains four 
triangles and every triangle is shared by two tetrahedra, we also have the constraint
\begin{equation}
\label{id2}
N_2=2 N_3 .
\end{equation}
Both relations (\ref{id1}) and (\ref{id2}) are shared by Euclidean DT and standard CDT. 
In the latter we also have the foliation constraint $2 N_2^{sss}\! =\! N_3^{T_5}$, which expresses the fact that in 
CDT every spacelike triangle is shared by two tetrahedra of type $T_5$, while every such tetrahedron contains 
exactly one spacelike triangle. In the present case, the analogous relation is
\begin{equation}
\label{id3}
2 N_2^{sss}=2 N_3^{T_2}+N_3^{T_3}+N_3^{T_5} .
\end{equation}
This is easily understood by counting all spacelike faces in the triangulation -- the right-hand side of (\ref{id3}) --
and noting that the number of spacelike triangles is half the number of spacelike faces.

In CDT we have two more constraints which explicitly involve the leaves of the preferred foliation.
One is the Euler constraint $N_0-N_1^s+N_2^{sss}= t_{tot} \tilde{\chi}$, where $t_{tot}$ counts the number of leaves
(for periodic boundary conditions in time), and $\tilde{\chi}$ is the Euler characteristic of a spatial section.
This constraint no longer exists in the generalized model, since we have Monte Carlo moves which change the quantity
$N_0-N_1^s+N_2^{sss}$. Furthermore, in standard CDT every spacelike triangle has three spacelike links and every 
spacelike link is shared by two spacelike triangles, yielding the relation $N_1^s\! =\! 3 N_2^{sss}/2$. 
As shown in \cite{thesis}, this can be generalized to the case at hand, leading to the linear relation
\begin{equation}
N_1^s=\frac{1}{2}(3N_2^{sss}-N_3^{T_2}) .
\end{equation}
Lastly, a constraint which does not have a counterpart in foliated CDT follows directly from our earlier observation of
$T_2$- and $T_3$-tetrahedra forming closed rings (assuming compact spatial topology), namely,
\begin{equation}
N_2^{sst}=N_3^{T_2}+N_3^{T_3} .
\end{equation}

We have checked that the Monte Carlo moves for the nonfoliated CDT model, described in Sec.\ \ref{sec:numsetup} below, 
are not compatible with the existence of other linear relations among the counting variables (\ref{countvar}). 
This means that we have a total of 5 such relations for 10 variables, compared with 5 relations for 7 counting variables
for standard CDT quantum gravity in 2+1 dimensions. In the next section, we will express the gravitational action
as function of the five remaining independent counting variables.

\section{Action and Wick rotation}
\label{sec:actions}

The gravitational path integral (\ref{eq:pigrav}) assigns to every spacetime geometry $[g]$ a complex amplitude $\exp(iS[g])$,
where $S[g]$ is its classical action. As already noted, we will use the same Regge implementation of the Einstein-Hilbert action
in 2+1 dimensions as previous work on CDT quantum gravity \cite{3d4d}, namely,
\begin{align}
\label{eq:reggeaction3d}
S_{\mathrm{Regge}}\!=\! k\!\!\! \sum_{\substack{\mathrm{spacelike}\\  l}} \!\!\! V(l)\frac{1}{i}
\left(2\pi- \!\!\!\!\!  \sum_{\substack{\mathrm{tetrahedra}\\ \mathrm{at}\, l}} \!\!\!\! \Theta\right)+
k\!\!\! \sum_{\substack{\mathrm{timelike}\\  l}} \!\!\! V (l)\left(2\pi-\!\!\!\!\!\sum_{\substack{\mathrm{tetrahedra}\\ \mathrm{at}\, l}}
\!\!\!\! \Theta\right)
\! -\!\lambda\!\!\!\!\!\! \sum_{\substack{ \mathrm{tetrahedra}\, T }}\!\!\!\!\!\! V(T),
\end{align}
where $k$ and $\lambda$ are the gravitational and cosmological couplings (up to rescaling), $V(l)$ and $V(T)$ the
volumes of a link $l$ and a tetrahedron $T$, and $\sum\Theta$ denotes the sum over dihedral angles contributed by
the tetrahedra sharing a link $l$. It was shown in
\cite{3d4d} that an analytic continuation $\alpha\!\mapsto\! -\alpha$ in the asymmetry parameter through the lower-half 
complex plane defines a nonperturbative Wick rotation which converts the amplitudes $\exp(iS_{\mathrm{Regge}})$ to 
real weights $\exp(-S_{\mathrm{Regge}}^\mathrm{eucl})$, and thereby makes it possible to analyze the path integral
with the help of Monte Carlo simulations. Maintaining the relation $\ell_t^2\! =\! - \alpha \ell_s^2$
between time- and spacelike length assignments, this implies that timelike links acquire a {\it positive} squared
length after the Wick rotation and therefore effectively become spacelike. The requirement that the full set of link
lengths correspond to a proper triangulation {\it after} the Wick rotation means that they have to obey triangle inequalities,
which in turn puts a restriction on the value of $\alpha$ before the Wick rotation, which for CDT in 2+1 dimensions takes 
the form $\alpha\! >\! 1/2$. 

Let us study how the enlargement of the ensemble of configurations in nonfoliated CDT affects the behaviour of
the Regge action (\ref{eq:reggeaction3d}) under the map $\alpha\!\mapsto\! -\alpha$. 
In the first term, $V(l)\! =\! 1$, because the link is spacelike. The plane orthogonal to the link has 
Lorentzian signature. Because we have imposed link causality, we will cross the light cone four times when
circling around the link in the plane. According to our complex angle prescription, each crossing adds
a real contribution $\pi/2$ to the total dihedral angle, such that the deficit angle -- the expression 
$(2\pi\! -\! \sum\Theta)$ inside the parentheses -- becomes purely imaginary, like
in usual CDT, and under the analytic continuation becomes a real deficit angle.

\begin{table}[t]
\begin{center}
\renewcommand{\arraystretch}{1.4}
\begin{tabular}{|c|c|c|}
\hline
tetrahedron & volume &  Wick rotation condition\\
\hline
\hline
$T_2$ & $\frac{1}{12}\sqrt{\alpha(3+\alpha)}$ & $0 < \alpha < 3$ \\
\hline
$T_3$ & $\frac{1}{12}\sqrt{1+4\alpha+\alpha^2}$ & $2-\sqrt{3} < \alpha < 2+\sqrt{3}$\\
\hline
$T_5$ & $\frac{1}{12}\sqrt{1+3\alpha}$ & $\frac{1}{3} < \alpha $ \\
\hline
$T_9$ & $\frac{1}{6}\sqrt{\frac{1}{2}+\alpha}$ & $ \frac{1}{2} < \alpha$ \\
\hline
\end{tabular}
\end{center}
\caption{Volumes of the four elementary tetrahedra and conditions on the asymmetry parameter $\alpha$, 
which ensure that the building blocks after the Wick rotation are well defined.}
\label{tab:tetravolumes}
\end{table}

In the second term, the plane orthogonal to the timelike link is Euclidean and remains so after the Wick rotation.
On the other hand, we have $V(l)\! =\!\sqrt{\alpha}$, which acquires a factor of $-i$ under the analytic continuation. 
This implies that the second term changes from real to purely imaginary, as it should.
To evaluate the third term we need the volumes of the tetrahedra, which are shown in Table\ \ref{tab:tetravolumes}. 
The three-volumes as functions of $\alpha$ are useful quantities to look at. In the Lorentzian sector $(\alpha \! >\! 0)$,
they are all real and positive. After Wick rotation, a vanishing of the volume $V(T)$ signals a geometric degeneracy 
of the underlying (Euclidean) tetrahedron $T$, 
associated with a violation of the triangle inequalities. In addition, note that for the Wick-rotated expressions to give the
correct contributions to the Euclidean action, the arguments of the square roots in the second column have to be negative
after the substitution $\alpha$ by $-\alpha$, leading to the restrictions on the original $\alpha$-values displayed in
the third column of the table. Since all of these constraints have to be satisfied simultaneously, we conclude that
in nonfoliated CDT in three dimensions we need
\begin{equation}
\frac{1}{2} < \alpha < 3
\end{equation}
in order for the usual Wick rotation to be well defined, which
is stronger than the corresponding condition $1/2 < \alpha$ in CDT, where only the building blocks $T_5$ and
$T_9$ are used.

Evaluating the Regge action (\ref{eq:reggeaction3d}) with the help of the expressions in Tables\ \ref{tab:tetraangles} 
and \ref{tab:tetravolumes}, and applying the Wick rotation leads to 
\begin{equation}
\label{eq:linearaction2}
S^\mathrm{eucl}=\widetilde{c_1} N_0 + \widetilde{c_2} N_3 + \widetilde{c_3} N_3^{T_2} + 
\widetilde{c_4} N_3^{T_3} + \widetilde{c_5} N_3^{T_5}
\end{equation}
for the Euclideanized Regge action, as function of a specific linearly independent subset of the counting variables 
(\ref{countvar}),
where the explicit functional form of the coefficients $\widetilde{c_i}=\widetilde{c_i}(k,\lambda,\alpha)$ has been
derived in \cite{thesis}.

For the special case $\alpha\! =\! -1$ and after some manipulations using the kinematical constraints, 
the action (\ref{eq:linearaction2}) can be written as
\begin{equation}
\label{eq:action_edt}
S^\mathrm{eucl}=-2 k \pi N_1 + \left(\frac{\lambda}{6\sqrt{2}}+6 k \arccos\frac{1}{3}\right)N_3,
\end{equation}
which coincides with the action of Euclidean dynamically triangulated gravity in three dimensions. 
We have also checked that by setting $N_3^{T_2}\! =\! N_3^{T_3}\! =\! 0$ the action (\ref{eq:linearaction2}) 
can be rewritten to precisely match the Regge action for standard CDT quantum gravity in 2+1 dimensions
given in \cite{3d4d}.

\section{Numerical set-up}
\label{sec:numsetup}
\subsubsection*{Monte Carlo moves}

\begin{figure}[t]
\centerline{\scalebox{0.75}{\includegraphics{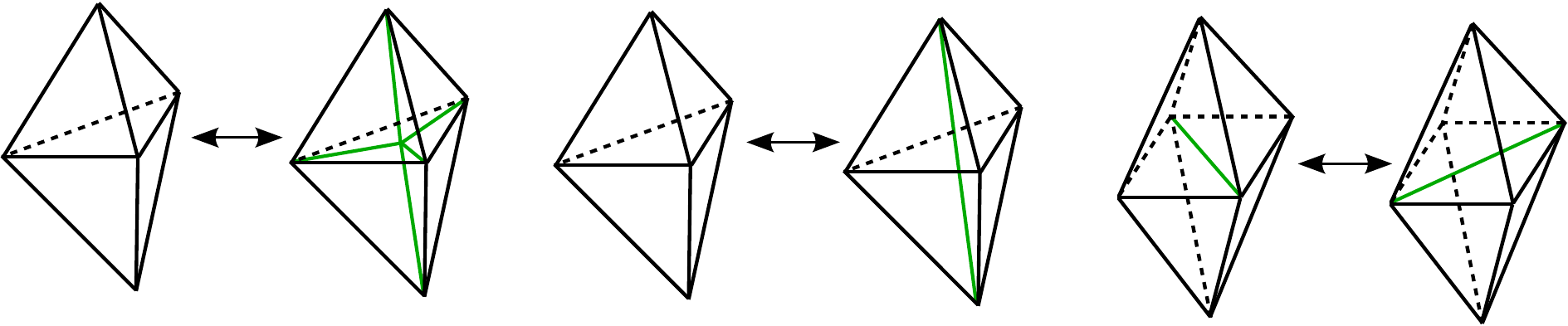}}}
\caption{The three generalized Pachner moves for CDT in 2+1 dimensions: 2-6 move (left), 2-3 move (centre) 
and 4-4 move (right). The links shown in green are removed or created by the corresponding move.}
\label{fig:genpach}
\end{figure}

To set up numerical simulations of nonfoliated CDT quantum gravity in 2+1 dimensions, we need to define a set of 
Monte Carlo moves. In this section, we will present a compact description of the moves, which fall into two groups;
further details can be found in \cite{thesis}. 
The first group contains generalizations of the moves that were already used for CDT simulations in 2+1 dimensions
\cite{3d4d}, and which in turn are adapted versions of the original Pachner moves for Euclidean DT \cite{pachner,gross}. 

Fig.\ \ref{fig:genpach} shows the three adapted Pachner moves in 2+1 dimensions. They all change the interior of 
a small compact region of the simplicial manifold, while leaving its boundary invariant. 
For CDT triangulations, once the location of a link to be added has been fixed, its type (timelike or spacelike) is also fixed.
This is no longer true in the nonfoliated CDT model, where each of these moves comes in several ``flavours". 
A move of this kind is called a ``$m$-$n$ move'' if $m$ and $n$ are the numbers of tetrahedra in the local simplicial 
neighbourhood before and after the move is executed. The 2-6 move is the only generalized Pachner move which creates 
a new vertex.

\begin{figure}[t]
\centerline{\scalebox{0.65}{\includegraphics{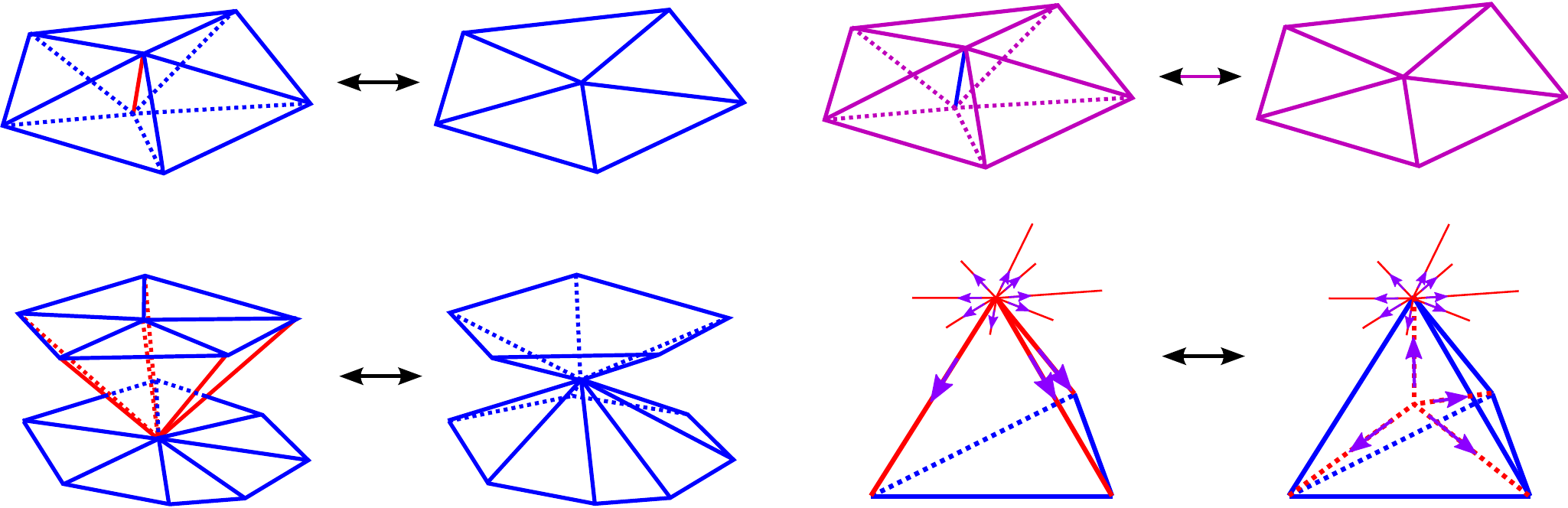}}}
\caption{New Monte Carlo moves in nonfoliated CDT. Three link collapse moves: bubble move (top left), 
pinching move (bottom left) and link collapse move for spacelike links (top right). Also shown is the
polar move (bottom right), with arrows indicating a time orientation. -- Spacelike links are blue, 
timelike ones red, and purple links can be either space- or timelike.}
\label{fig:newmoves}
\end{figure}

The Monte Carlo moves in the second group are new compared to standard CDT. Three of them
implement the collapse of a link, and only differ in the types of links and the local neighbourhood involved,
as illustrated by Fig.\ \ref{fig:newmoves}.
They can be seen as special cases of the most general link collapse move, of which we currently do not know
whether and how it can be implemented efficiently. 

The bubble move (Fig.\ \ref{fig:newmoves}, top left) operates on a ring of $T_2$-tetrahedra with a single timelike link 
in its interior, forming a ``bubble" according to the definition given in Sec.\ \ref{sec:nfct21}. 
It collapses the timelike link to a single vertex and simultaneously collapses the bubble to a spatial disc. 
The pinching move (Fig.\ \ref{fig:newmoves}, bottom left) operates on a pair of spatial discs whose centres are connected 
by a timelike link. It collapses this link, leading to a configuration where the discs touch in a single vertex, thereby
forming a ``pinching" as described in Sec.\ \ref{sec:nfct21} (c.f. Fig.\ \ref{fig:rings}d).

We have also implemented a move which collapses a spacelike link (Fig.\ \ref{fig:newmoves}, top right). 
Note that in the configuration before the collapse the link types of the upper and lower disc do not necessarily 
have to match. In the special cases when they do, we call this move {\it symmetric}. It means that during the
collapse, only links of the same type get identified pairwise. To keep the complexity of the implementation at 
a manageable level, we have restricted ourselves to the symmetric version of this move.

Lastly, recall our introduction in Sec.\ \ref{sec:ncdt_2d} of an isolated source and sink of time in 1+1 dimensions 
(Fig.\ \ref{fig:sourcesink2d}). We will use a straightforward generalization to 2+1 dimensions 
of these local (causality-violating) configurations as our boundary conditions. The polar move operates on the 
neighbourhood of such a source or sink of time, and moves it around. Fig.\ \ref{fig:newmoves} (bottom, right) illustrates the 
situation for a time source, initially located at the top of the single tetrahedron on the left. 
The move subdivides the tetrahedron into four, with the newly created vertex at the centre becoming the 
new source of time.

An important feature of a set of Monte Carlo moves is that it should be ergodic, that is, any element of the configuration
space can be reached in a finite number of moves. In our case, the configuration space 
$\tilde{\mathcal{C}}$ consists of all locally causal gluings of the elementary building blocks $T_2$, $T_3$, $T_5$ and $T_9$
that can be time-oriented consistently, and satisfy further
regularity conditions specified in the next subsection. We have made the standard
choice of a direct product $[0,1]\!\times\! S^2$ for the spacetime topology. 
It is possible that the moves described here are ergodic in this configuration space; in fact,
the original motivation for introducing additional building blocks was to let us move around in the
space of triangulations more efficiently. However, we do not have a proof of ergodicity, and suspect this could be
rather nontrivial, given the nonlocal character of part of the causality conditions.

\subsubsection*{Defining the ensemble}

As already mentioned in Sec.\ \ref{cdt}, previous simulations of CDT in 2+1 dimensions have worked with a fixed
spacetime topology of direct-product form $[0,1]\!\times\! {}^{(2)}\Sigma$, or $S^1\!\times\! {}^{(2)}\Sigma$ if time is
compactified. The standard, simplest choice\footnote{For a recent investigation with toroidal slices, see 
\cite{bl}.} for the spatial topology -- which we will also employ in the present work -- is the sphere, ${}^{(2)}\Sigma\! =\! S^2$.
A posteriori, the choice of compactifying time in this case does not appear to make much of a difference, because it
turns out that the {\it dynamics} of 2+1 CDT quantum gravity (for sufficiently large time extension $t_{tot}$ of the
configurations) drives the shape of the universe towards a de Sitter space with $S^3$-topology \cite{3dcdt}. 

As we will now go on to explain, the most convenient choice of boundary conditions for nonfoliated CDT
is that of a direct-product spacetime $[0,1] \times S^2$, where the beginning and end of time are allowed
to degenerate to a point, 
leading effectively to an $S^3$-topology. Recall that
in simulations of CDT quantum gravity, the number of time steps $t_{tot}$ is fixed. Since genuine foliated
CDT triangulations form a subset of the present ensemble $\tilde{\cal C}$, the question arises whether it is possible to 
go from one strictly foliated configuration to another one with a different number $t'_{tot}$ of time steps via 
nonfoliated configurations and using the Monte Carlo moves described in the last subsection. 
As explained in detail in \cite{thesis},
the answer is yes. It follows that if there is a region in the phase diagram where the configurations are close to foliated, 
the standard CDT notion of the ``number of time steps" will also make sense approximately 
and one can ask which equilibrium value for this quantity is found after thermalization.
During early test simulations in the ensemble $\tilde{\cal C}$ with compactified time we did find configurations that 
were approximately foliated, but the number of time steps would not thermalize properly. 
We have been able to circumvent this problem by not compactifying time, and adding a source and a sink of time 
as the two poles of a three-sphere.

Another technical issue which appeared during early test runs was that the simulations would often end up 
in ``frozen" states where virtually no progress could be made using the implemented Monte Carlo moves. 
The problem could be traced back to the presence of globally self-overlapping bubbles, winding once or
multiple times around the spatial sphere (c.f. our discussion in Sec.\ref{sec:nfct21}).
Since we were unable to overcome this problem by finding additional moves, we looked for a mechanism to 
prevent the globally self-overlapping bubbles from appearing. 
We found that these problematic structures do not form when we forbid all moves which merge or split bubbles. 
The moves of Fig.\ \ref{fig:newmoves} are essentially unaffected by these restrictions (see \cite{thesis} for details).
The simulations on the reduced ensemble behaved much better after this alteration, although 
there are still phase space regions where they do not thermalize sufficiently well, as we will discuss later.

Finally, we will use local regularity conditions for the gluings that make the triangulations into
simplicial manifolds, which means that each (interior) vertex has a ball-like neighbourhood whose surface is a triangulated
two-sphere. This is the choice made in
most of the work on higher-dimensional CDT quantum gravity, and will allow for a better comparison of results. 

\subsubsection*{Re-introducing time}

The distribution of spatial volume as a function of time is an important large-scale observable, and its analysis has been 
instrumental in relating CDT quantum gravity to a de Sitter minisuperspace cosmology in 2+1 and 3+1 
dimensions \cite{3dcdt,desitter}. In order to perform a similar analysis also in nonfoliated CDT, we need to define a time 
coordinate on its generalized configurations. As explained in the introduction, the fact that spatial slices will generally
branch and form ``bubbles" means that we can no longer use them to define a distinguished time variable. 

To explain our alternative prescription of ``time", consider a time-oriented member of the configuration space 
$\tilde{\cal C}$.
Given a vertex $V$, consider the set of all future-oriented paths connecting $V$ with the 
north pole. The number of links in each path defines a distance between $V$ and the north pole. By averaging
this quantity over all paths we obtain an average distance $d_f$. Repeating the procedure for past-oriented paths, connecting
$v$ to the south pole, gives another average distance $d_p$. The time coordinate of $V$ is then defined as 
$t\! =\! d_f-d_p$. Note that for foliated CDT configurations, this coincides with the usual discrete proper time, up to a trivial factor.
We have experimented with other notions of time, including that of shortest distance to the poles; they generally lead 
to a ``washing out" of the tetrahedron distributions described below. It is possible that alternative notions of 
time are more appropriate or practical for observables different from the ones studied here.

Since the number of oriented paths between a vertex and a pole can become very large, in practice we 
used a modified algorithm, which calculates $t$ in an approximate fashion. 
For each vertex we constructed a fixed number of future-oriented paths, using a random process which jumps 
iteratively from a vertex to a randomly chosen future neighbour until the north pole is reached. 
We then repeated this process for past-oriented paths and finally calculated the time coordinate as the 
difference of both average distances. 

Given this new notion of time, we can now also assign an (average) time to spatial slices.
We define a {\it spatial slice} in nonfoliated CDT -- with boundary conditions as specified above --
as any subset of spatial triangles forming a two-sphere, such that by cutting along the 
sphere the spacetime triangulation decomposes into two disconnected parts, with time flowing consistently from
one side of the cut to the other. In other words, the future-pointing arrows introduced in Fig.\ \ref{fig:flowoftime3d} 
are all lined up to point in the same direction away from the slice.
The time coordinate we assign to such a slice is the average of all time coordinates of its vertices. Note that
unlike in standard CDT, where different spatial slices are always disjoint, spatial slices here can have any amount
of overlap. 

We now have all the ingredients to measure the desired volume profiles. 
Since the number of spatial slices of an individual path integral configuration can become very large, we use a 
statistical method to generate a subset of spatial slices which is evenly distributed along the time direction. 
In order to perform an ensemble average of the volume distribution we use a nontrivial averaging 
algorithm, details of which are described in \cite{thesis}.

\section{Exploring the phase diagram}
\label{sec:explo}

We have developed the necessary Monte Carlo simulation software, using C{}\verb!++! as programming language and 
taking advantage of object-oriented design principles to incorporate modularity and flexibility into the software. 
We had anticipated that the software would be more complex than for the CDT simulations, but in the event its 
complexity even surpassed our expectations. An extended discussion of the details of the software implementation, 
with special emphasis on the Monte Carlo moves, is given in \cite{thesis}. In what follows, we will present the
results obtained with the simulation software, beginning with an exploration of the phase diagram of 
nonfoliated CDT quantum gravity in 2+1 dimensions.

The Regge form (\ref{eq:reggeaction3d}) of the gravitational action contains two couplings, $k$ and $\lambda$, 
which are proportional to the inverse bare Newton's constant and the bare cosmological constant respectively. 
When evaluating the action on causal triangulations, a third parameter -- the asymmetry $\alpha$ -- naturally 
appears because of the distinction between space- and timelike links. 
Together they span a three-dimensional space of bare actions. 
Of course, from the way $\alpha$ is introduced in the regularized theory, there is no a priori reason why it 
should play the role of a coupling constant. Different $\alpha$-values should lead to the same continuum
gravity theory. This expectation is consistent with the dynamical results found below.\footnote{In
3+1 dimensions the status of the analogous parameter $\alpha$ is more involved \cite{trans,physrep}.}

\begin{figure}[t]
\centerline{\scalebox{0.8}{\includegraphics{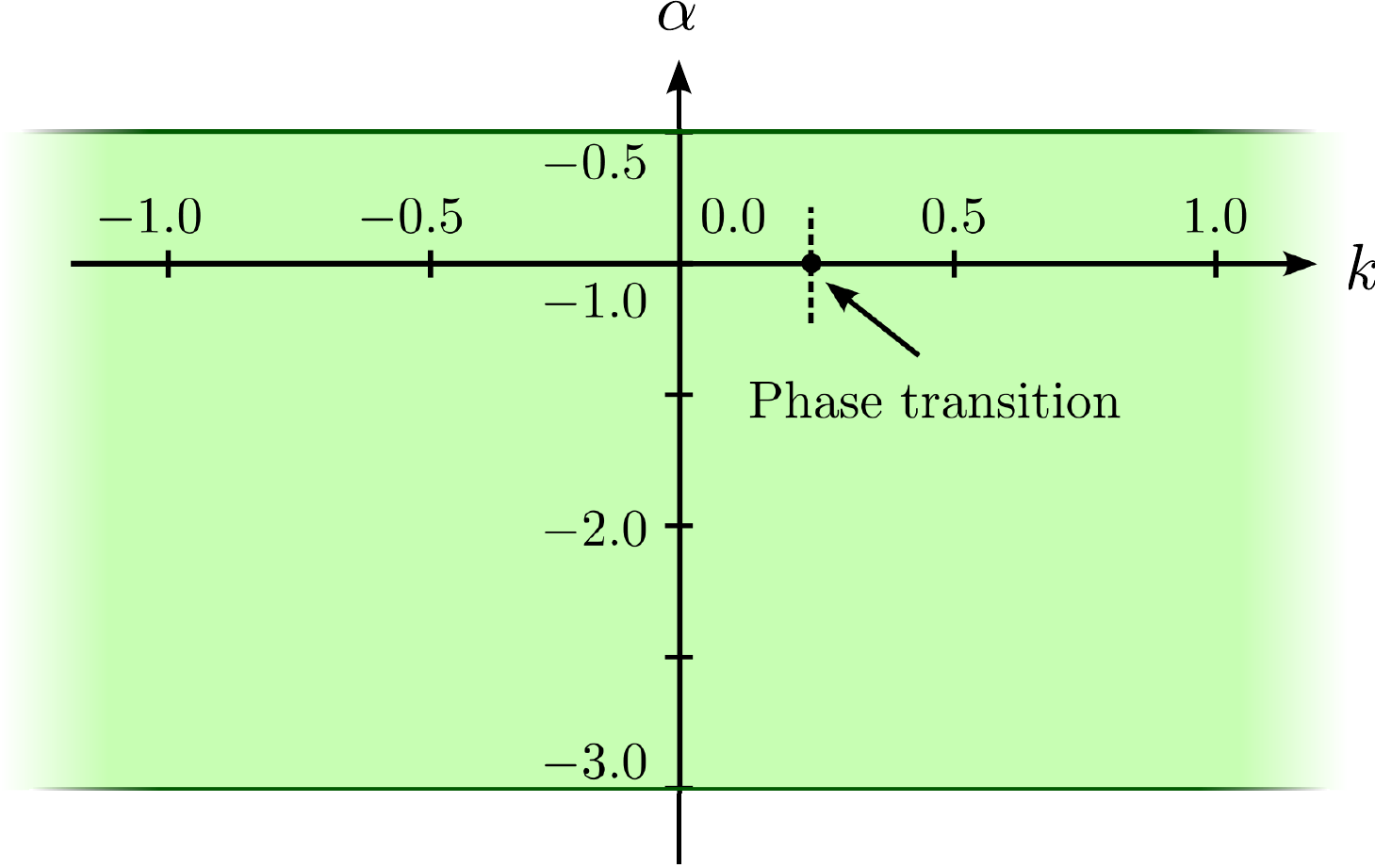}}}
\caption{The phase diagram of nonfoliated CDT quantum gravity is the region inside the 
strip $-3\! <\!  \alpha\!  <\!  -1/2$, visualized in green. 
Our investigation has probed the phase space along the two axes drawn in the figure: at various $k$-values for 
constant $\alpha\! =\! -1$,
and at various $\alpha$-values for constant $k\! =\! 0$.}
\label{fig:pd_lcdt}
\end{figure}

As usual in dynamically triangulated systems, we do simulations at fixed system size $N_3$ and then perform
a finite-size scaling analysis to extrapolate to the limit of infinite size. This
means that the phase diagram of the model is spanned by the parameters $k$ and $\alpha$.
As we have derived in Sec.\ \ref{sec:actions}, the existence of a Wick rotation limits the allowed values for $\alpha$ 
to the region $1/2\!  <\!  \alpha\! <\!  3$. Sticking with the notation ``$\alpha$" for this parameter also after the
analytic continuation, the Wick rotation maps this region to the range $-3\! <\!  \alpha\!  <\!  -1/2$. 
The phase diagram of our generalized CDT model is therefore a strip bounded by these two $\alpha$-values,
as illustrated by Fig.\ \ref{fig:pd_lcdt}.

In the following we will analyze the dynamics of our model at a range of points along the two axes drawn in
the figure.
While the simulations work well on the axis defined by constant $\alpha\! =\! -1$, 
we encountered difficulties when exploring the axis of constant $k\! =\! 0$ in the region $-3\! <\! \alpha\! <\! -1$. 
As one moves away from $\alpha\! =\! -1$ towards $\alpha\! =\! -3$, the number of accepted Monte Carlo moves goes
down significantly and the thermalization time increases rapidly. 
A closer analysis revealed that the severity of the problems correlates with the presence of bubbles with a
complicated internal structure. 
These problems imply that we currently must concentrate our investigation on the region $-1\! <\! \alpha\! <\! -1/2$.

\subsubsection*{Bounds on the vertex density}

In 2+1 dimensions there are kinematical bounds on the ratios of certain counting variables, like
the vertex density $N_0/N_3$ and the link density $N_1/N_3$. 
In the case of CDT the link density satisfies $1\! \leq\! N_1/N_3\!\leq\! 5/4$ \cite{3d4d}, which should be compared 
with the weaker bound $1\!\leq\! N_1/N_3 \!\leq\! 4/3$ for DT.
Using the linear relations (\ref{id1}) and (\ref{id2}), one easily derives $N_0/N_3=N_1/N_3-1$ in the 
infinite-volume limit, which means that we can translate the 
link density bounds into the vertex density bounds 
$0\!\leq\! N_0/N_3\!\leq\! 1/4$ for CDT and $0\! \leq\! N_0/N_3 \!\leq\! 1/3$ for Euclidean triangulations.

The derivation of the link density bound in CDT involves the spatial Euler constraint, which is not present in the 
ensemble $\tilde{\cal C}$ we have specified in Sec.\ \ref{sec:numsetup} above.
To find the analogous bound for nonfoliated CDT configurations we follow \cite{3d4d} 
in considering all Monte Carlo moves that create a vertex. 
Since all of them change the number of tetrahedra by some amount $\Delta N_3$, the strategy is to select
those moves for which $\Delta N_3$ is minimal. 
Starting with a minimal triangulation and repeatedly applying only the selected moves, the vertex density 
and thus also the link density will be maximized, and the corresponding bounds follow upon taking the infinite-volume limit.

\begin{figure}[t]
\centerline{\scalebox{0.7}{\includegraphics{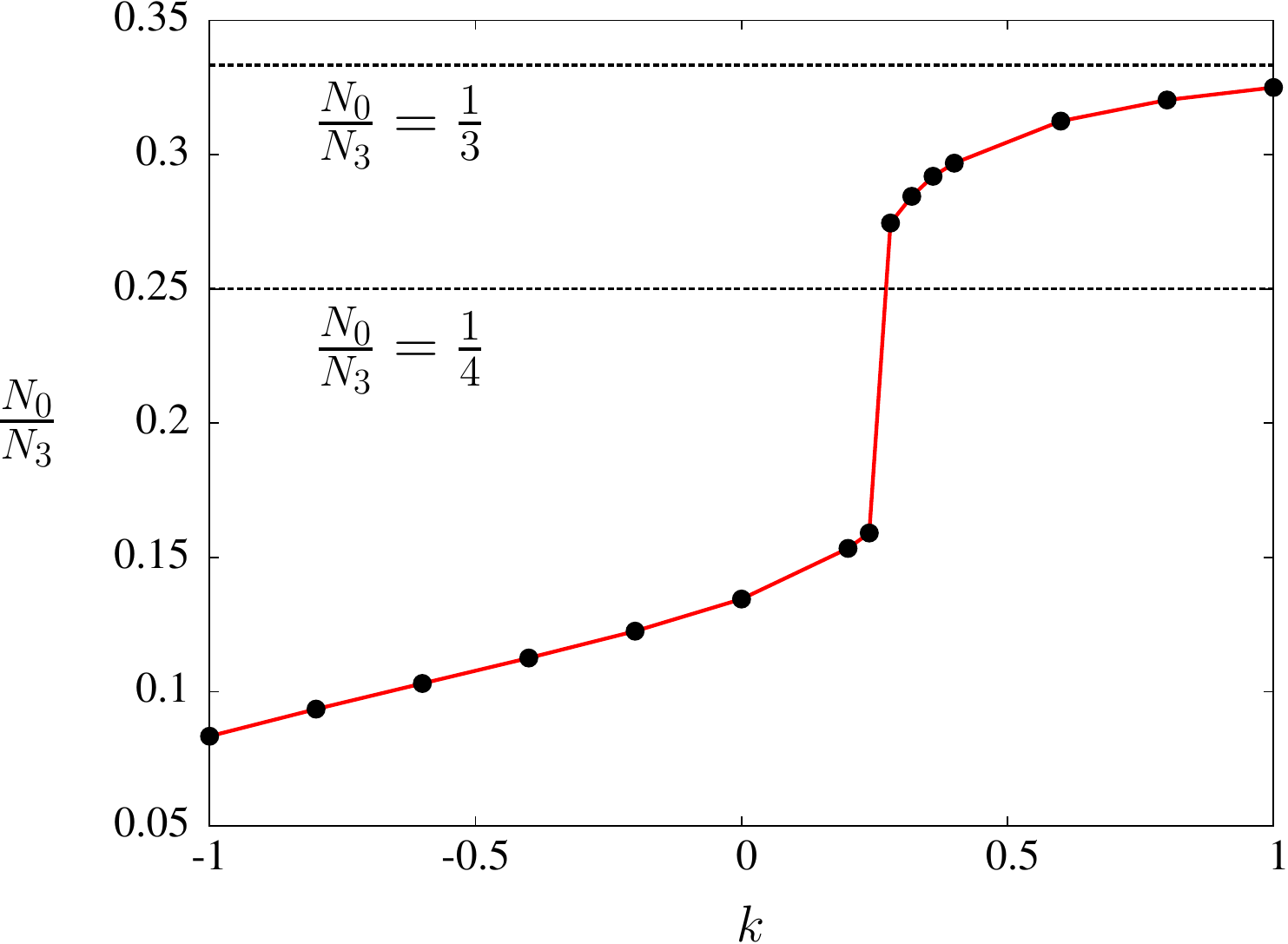}}}
\caption{Measurement of the average vertex density $N_0/N_3$ as function of the inverse gravitational coupling $k$, 
for  $\alpha\! =\! -1$ and system size $N_3\! =\! 40.000$. 
The dots represent actual measurements, the lines linear interpolations. 
The dashed lines mark the kinematical bounds for standard DT (upper line) and CDT (lower line).}
\label{fig:n0n3_k}
\end{figure}

In the case at hand we have two Monte Carlo moves which create one vertex and three tetrahedra, namely, the 
bubble move and the polar move described in Sec.\ \ref{sec:numsetup} above. Both are unconstrained moves 
which can always be executed. We conclude that in nonfoliated CDT quantum gravity the vertex and link densities 
satisfy the bounds
\begin{equation}
0\leq \frac{N_0}{N_3} \leq \frac{1}{3} \qquad \mathrm{and} \qquad
1\leq \frac{N_1}{N_3} \leq \frac{4}{3} .
\end{equation}
These relations agree with those for Euclidean DT, but 
the configurations which saturate them differ substantially, as we will see. 
Note that a relaxation (in the sense of \cite{thor}) of the local regularity conditions for a simplicial manifold
would weaken these bounds, since then bubbles with fewer than three tetrahedra can occur.

\begin{figure}[t]
\centerline{\scalebox{0.7}{\includegraphics{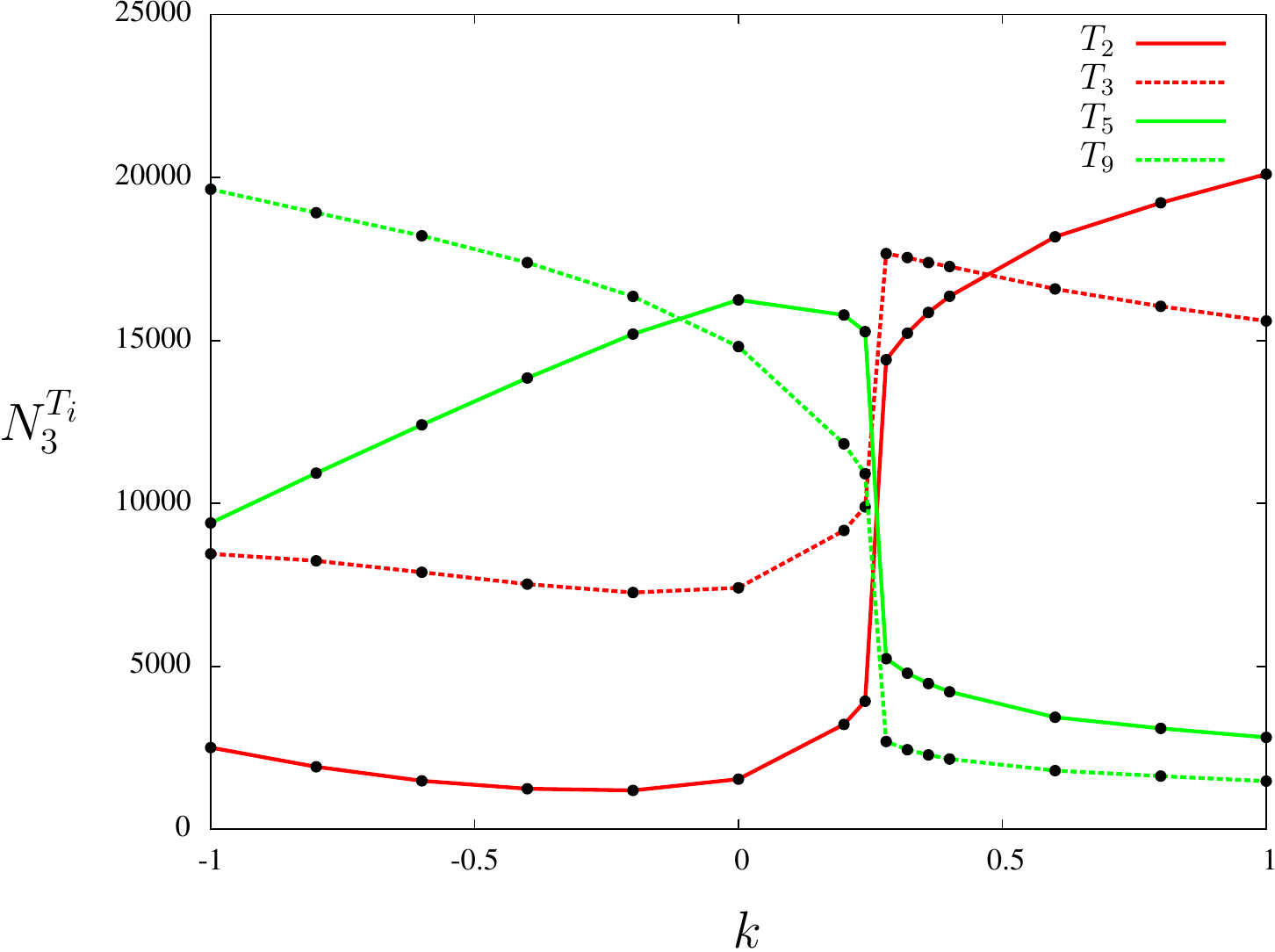}}}
\caption{The numbers of tetrahedra of each of the four types, averaged over the sampled triangulations, as function of the 
coupling $k$, at $\alpha\! =\! -1$ and $N_3\! =\! 40.000$.}
\label{fig:n3_k}
\end{figure}

Fig.\ \ref{fig:n0n3_k} shows the measurements of the average vertex densities for various values of the 
coupling $k$, from simulations with $\alpha\! =\! -1$ and system size $N_3\! =\! 40.000$. 
The vertex density increases monotonically with $k$, which is 
not surprising since (at fixed $N_3$) larger values of $k$ favour the creation of vertices. 
This can be seen easily by rewriting eq.\ (\ref{eq:action_edt}) with the help of the kinematical constraints, yielding 
$S^\mathrm{eucl}\! =\! -2 k \pi N_0$ plus a term proportional to $N_3$. 
As expected, the measured curve in Fig.\ \ref{fig:n0n3_k} 
approaches the upper kinematical bound of $1/3$ for large values of $k$. 
We also see a clear signal of a phase transition between $k\! =\! 0.24$ and $k\! =\! 0.28$, from a phase of low 
to one of high vertex density, reminiscent of the first-order transitions in the inverse gravitational coupling
found in both DT \cite{edt3d} and CDT \cite{3dcdt}. Analogous measurements for fixed $k\! =\! 0$ and varying 
$\alpha$ show that the vertex densities are approximately constant at low values, without any sign of a phase 
transition. Of course, since we are only investigating the region $-1\! <\!\alpha\! <\! -1/2$ of the phase diagram,
we cannot exclude the presence of further phase transitions in the complementary region.

\subsubsection*{Emergence of foliated triangulations}

Foliated CDT geometries form a subset of the ensemble $\tilde{\cal C}$, characterized by the condition 
$N_3^{T_2}\! =\! N_3^{T_3}\! =\! 0$.\footnote{With periodic boundary conditions in time one could in principle 
construct a triangulation obeying $N_3^{T_2}\! =\! N_3^{T_3}\! =\! 0$ consisting of a single bubble winding 
around both space and time, which clearly is not foliated. However, such configurations do not lie in $\tilde{\cal C}$
because in our case time is not compactified and we do not allow for globally self-overlapping bubbles.} 
By plotting the number of tetrahedra of these two types as function of the couplings we can therefore look for 
regions in phase space where foliated triangulations emerge dynamically. Fig.\ \ref{fig:n3_k} shows the numbers of 
all tetrahedral types used, averaged over the configurations sampled from $\tilde{\cal C}$, as function of the coupling 
constant $k$. In the phase with low vertex density on the left, although the building blocks of standard CDT dominate, 
also the other two types appear in significant numbers, from which we deduce that the triangulations 
along the line $\alpha\! =\! -1$ apparently are not foliated.

\begin{figure}[t]
\centerline{\scalebox{0.7}{\includegraphics{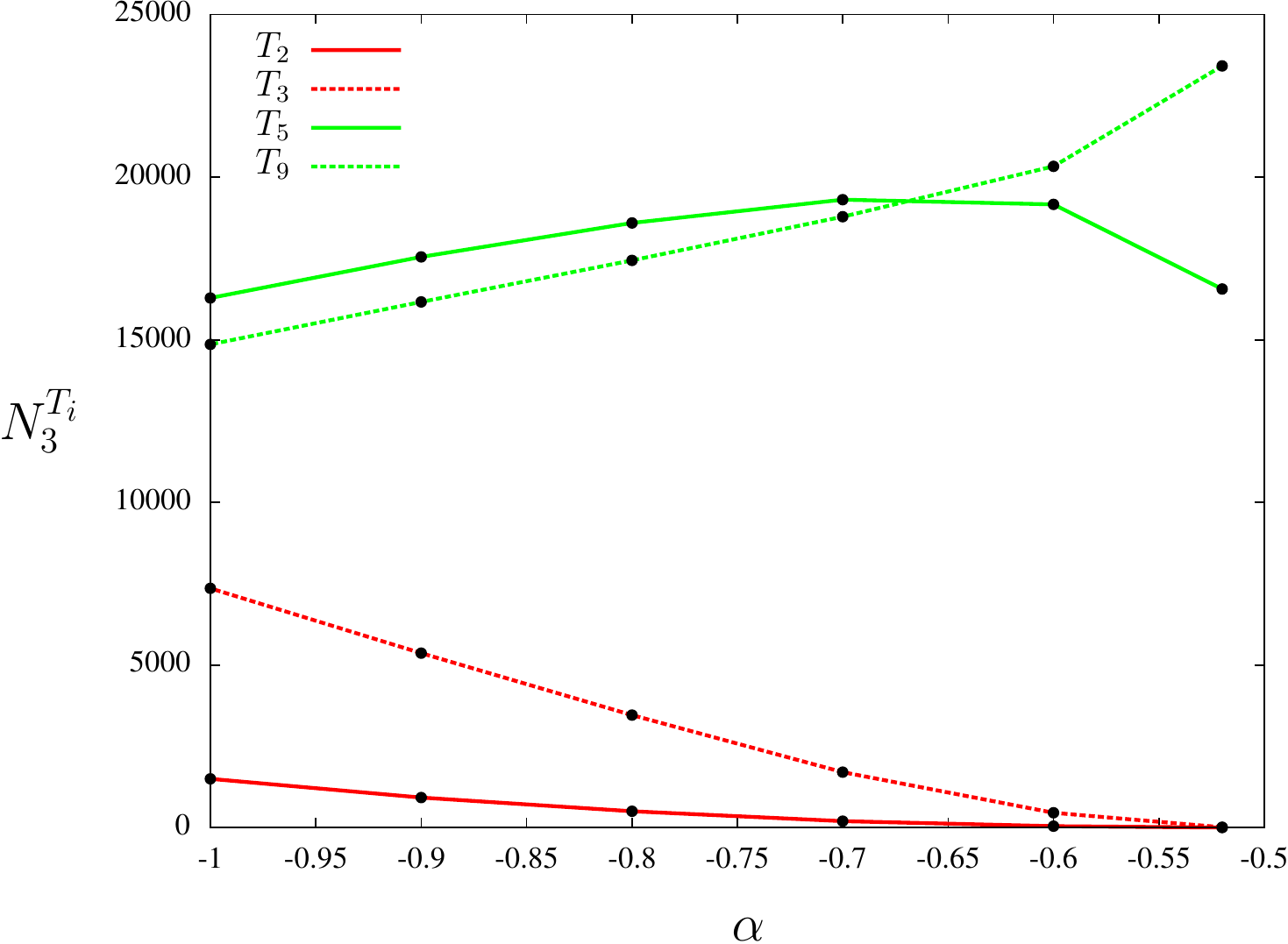}}}
\caption{The numbers of tetrahedra of each of the four types, averaged over the sampled triangulations, as function of the 
coupling $\alpha$, at $k\! =\! 0$ and $N_3\! =\! 40.000$.}
\label{fig:n3_alpha}
\end{figure}

Conversely, Fig.\ \ref{fig:n3_alpha} shows the expectation values of the numbers of tetrahedra at fixed $k\! =\! 0$, 
as function of $\alpha$. Note that the phase boundary $\alpha\! =\! -0.5$ does not belong to the phase diagram, 
since the Wick rotation is not defined there. 
The measurements corresponding to the rightmost data points in the figure have been performed at $\alpha\! =\! -0.52$. 
We find that both $N_3^{T_2}$ and $N_3^{T_3}$ approach zero as we move towards the phase boundary. 
At $\alpha\! =\! -0.52$ we have measured $\langle N_3^{T_2}\rangle\!\approx\! 2.9$ and 
$\langle N_3^{T_3}\rangle\!\approx\! 14.3$,
which means that in the entire system consisting of 40.000 tetrahedra almost none of the building blocks belong to 
the new types $T_2$ and $T_3$.

We conclude that the configurations appearing close to $\alpha\! =\! -0.5$ are almost perfectly foliated and belong to 
the phase with low vertex density. Effectively, the dynamics should therefore be very close to the known dynamics of 
2+1 dimensional CDT in the extended phase \cite{3dcdt}, and we would expect the geometries 
to be macroscopically extended with a characteristic blob-shaped volume distribution. 
These expectations will be confirmed later on.

\section{Tetrahedron distributions}
\label{sec:tetdist}

We have seen in the last section that foliated configurations emerge close to the boundary of the phase diagram. 
As we move away from the boundary, the configurations become less foliated. 
A strict foliation is attained whenever $N_3^{T_2}\! =\! N_3^{T_3}\! =\! 0$, but it is unclear how to translate nonzero values 
into a measure of foliatedness of a triangulation. 
We would like to have a more refined observable which tells us how foliated a triangulation is. 
In the following we will use tetrahedron distributions based on the time coordinate introduced in
Sec.\ \ref{sec:numsetup} as a qualitative tool to investigate the degree of foliatedness of a triangulation.

To start with, let us assume that all vertices have been assigned a time coordinate using the algorithm described in 
Sec.\ \ref{sec:numsetup}. For each tetrahedron we then calculate the sum of the time coordinates of its four vertices and 
round this value to the nearest integer. By definition, this value gets assigned to the tetrahedron 
as its new time coordinate. This ``tetrahedron time" clearly has a different relative normalization compared to the ``vertex time"
from which it was derived, but this does not matter as long as we do not use both time coordinates simultaneously.

\begin{figure}[t]
\centerline{\scalebox{1.0}{\includegraphics{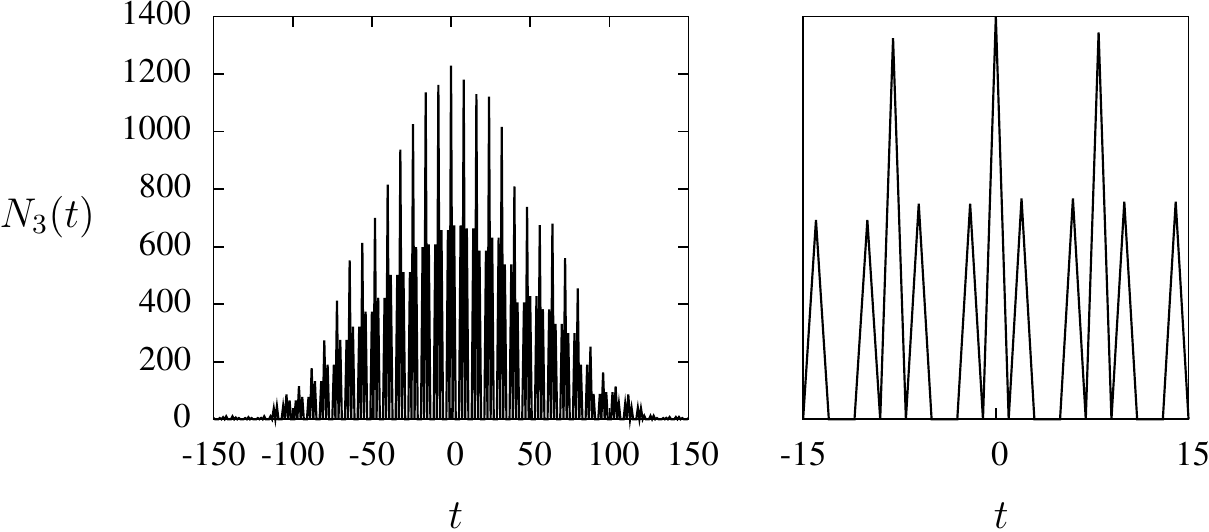}}}
\caption{Typical tetrahedron distribution of a strictly foliated CDT configuration as function of ``tetrahedron time",
extracted from a simulation 
with foliation constraint enabled (left). Zooming in on the central region of the distribution, one obtains the graph
shown on the right.}
\label{fig:prof_cdt}
\end{figure}

In a given configuration, we can now count the number of tetrahedra that share the same value of (tetrahedron) time,
and plot these numbers as a function of time to generate a tetrahedron distribution. 
Fig.\ \ref{fig:prof_cdt} (left) shows such a distribution for a strictly foliated CDT geometry, which we have generated 
by running our simulation with foliation constraint enabled. 
We observe that the distribution appears as a superposition of two blob-shaped distributions. 
One can show that one of them consists of $T_5$-tetrahedra and the other one of $T_9$-tetrahedra \cite{3dcdt}. 
An enlarged version of the central part of the distribution is shown in Fig.\ \ref{fig:prof_cdt} (right), which illustrates
that the peaks are organized in groups of three. One can show that every such group corresponds 
to a ``thick slice'', which is the triangulation enclosed between two adjacent simplicial spatial hypermanifolds \cite{thesis}, 
the higher-dimensional analogue of a ``strip" in 1+1 dimensions. Note that for pure CDT configurations 
the tetrahedron time is effectively a refinement of the number of time steps associated with the preferred foliation, 
similar to what was considered in \cite{semicl} to produce finer-grained volume distributions in 3+1 dimensions.

\begin{figure}[t]
\centerline{\scalebox{0.75}{\includegraphics{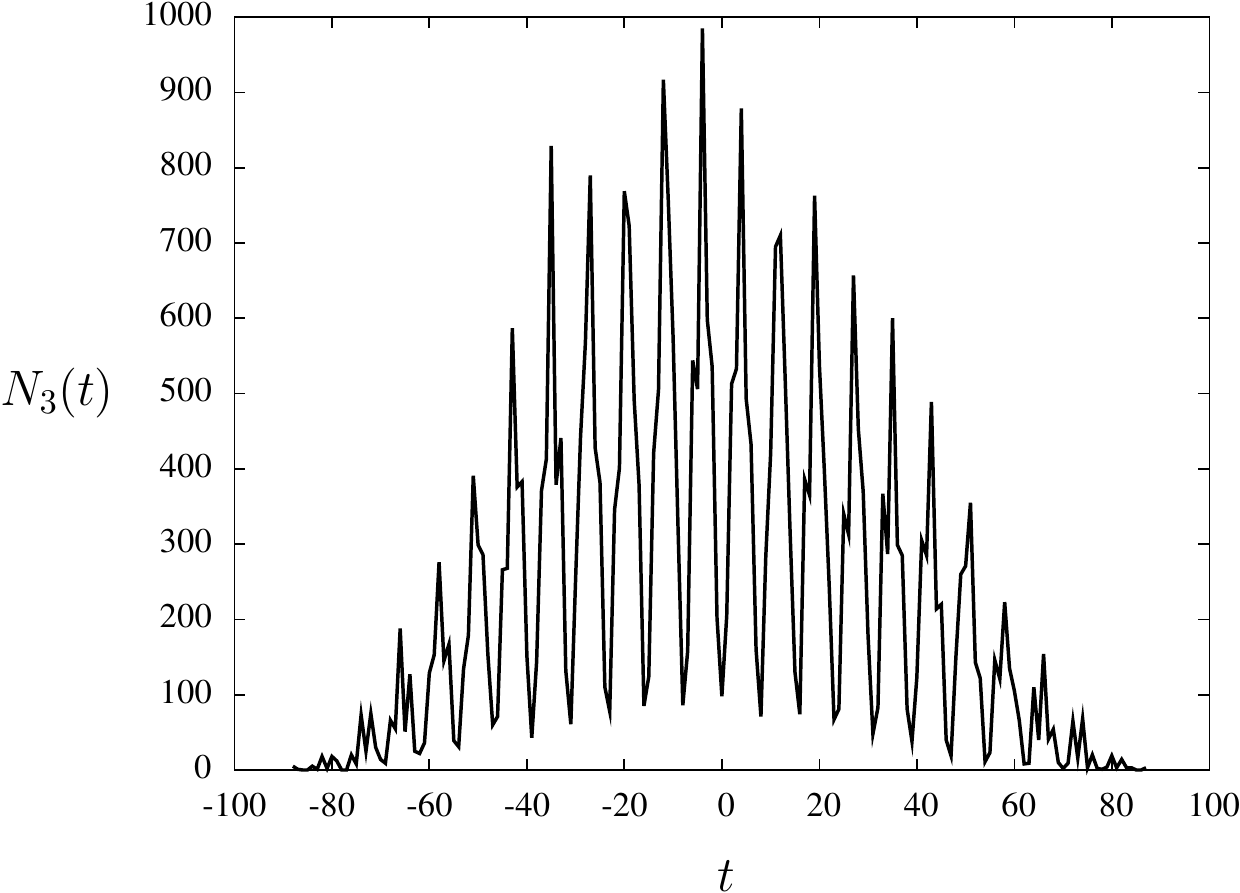}}}
\caption{Tetrahedron distribution of a typical path integral history from a simulation of 
the generalized CDT model at $k=0.0,\ \alpha=-0.7$.}
\label{fig:prof_k0.0a-0.7}
\end{figure}

Let us return to the more general setting of nonfoliated CDT quantum gravity and focus on triangulations which 
are a little further away from the phase boundary. Fig.\ \ref{fig:prof_k0.0a-0.7} shows the tetrahedron distribution 
of a single triangulation extracted from a simulation at $k\! =\! 0,\ \alpha\! =\! -0.7$, with volume $N_3=40.000$. 
We observe a sequence of peaks, with some remnants of the three-peak structure exhibited by Fig.\ \ref{fig:prof_cdt}. 
The tendency of these structures to become blurred most likely depends both on changes in the actual triangulation 
and on the precise algorithm used to define the time coordinate and the tetrahedron distribution. 
Consequently, the relevant information lies not so much in the precise structure of the peaks, but in the overall pattern 
formed by the succession of all the peaks. Comparison of the two configurations suggests that
every peak in Fig.\ \ref{fig:prof_k0.0a-0.7} corresponds to a group of three peaks in Fig.\ \ref{fig:prof_cdt} 
and describes a structure which resembles a thick slice in a foliated triangulation. 
Another aspect in which the two configurations differ is the fact that the gaps between each group of three peaks
in Fig.\ \ref{fig:prof_cdt} (right) -- marking the location of spatial triangulated hypermanifolds --  are getting filled in when the
foliation is relaxed. This can be interpreted as a ``decoration" of the spatial slices by the creation of bubbles. 

Based on these findings we can interpret the pattern shown in Fig.\ \ref{fig:prof_k0.0a-0.7} as a triangulation 
where decorated spatial slices alternate with modified thick slices. 
A triangulation exhibiting such a structure will be called \emph{weakly foliated}. This is obviously not a sharp definition 
since we are not providing a sharp criterion for when a weakly foliated triangulation changes into a truly nonfoliated one. 
We will take an operational point of view here and consider a triangulation to be weakly foliated whenever 
the tetrahedron distribution shows the characteristic alternating pattern.

\begin{figure}[t]
\centerline{\scalebox{1.0}{\includegraphics{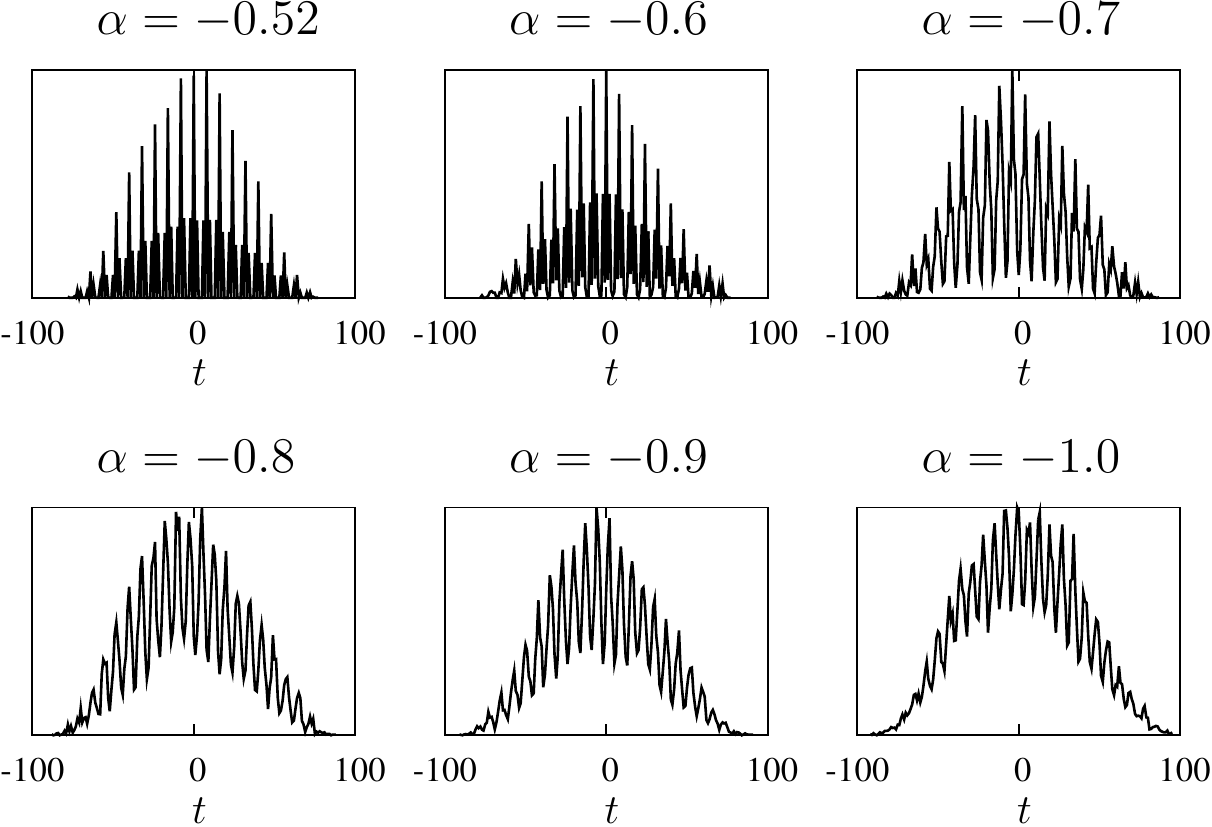}}}
\caption{Tetrahedron distributions of typical path integral configurations from simulations at coupling
$k=0.0$ and system size $N_3\! =\! 40.000$, for various choices of the asymmetry parameter $\alpha$. Note
that all geometries display some degree of being weakly foliated, which becomes weaker with increasing $| \alpha |$. }
\label{fig:prof_alpha}
\end{figure}

The next important step is to understand how the foliatedness of a triangulation changes as one moves 
around in phase space. Fig.\ \ref{fig:prof_alpha} shows a sequence of typical tetrahedron distributions of single 
triangulations extracted from simulations at $k\! =\! 0$, for various choices of $\alpha$. 
From an almost strict foliation at $\alpha\! =\! -0.52$ the signal -- although remaining distinctly visible -- gradually weakens as we 
move towards $\alpha\! =\! -1$. It would be interesting to follow the development of this pattern beyond this point
towards the other phase boundary, but technical issues currently prevent us from doing so, as we have discussed earlier.
We have performed a similar analysis on the line of constant $\alpha\! =\! -1$, and have observed
that the alternating pattern remains visible, but becomes less pronounced when one moves from $k\! =\! 0$ 
towards $k\! =\! -1$, indicating a further weakening of the foliation. 
When going from $k\! =\! 0$ in the other direction towards the phase transition, the data quality decreases significantly, 
to such an extent that an interpretation based on the tetrahedron distribution becomes unreliable. 
To summarize, it appears that all investigated configurations in the phase of 
low vertex density exhibit some kind of (weak) foliation, whose degree varies significantly, from 
an almost strict foliation near the phase boundary at $(k,\alpha)\! =\! (0,-0.5)$ to a much less pronounced one
for larger $| \alpha |$.

\section{Volume distributions}
\label{sec:voldist}

\subsubsection*{A phase of extended geometry}

\begin{figure}[t]
\centerline{\scalebox{1.0}{\includegraphics{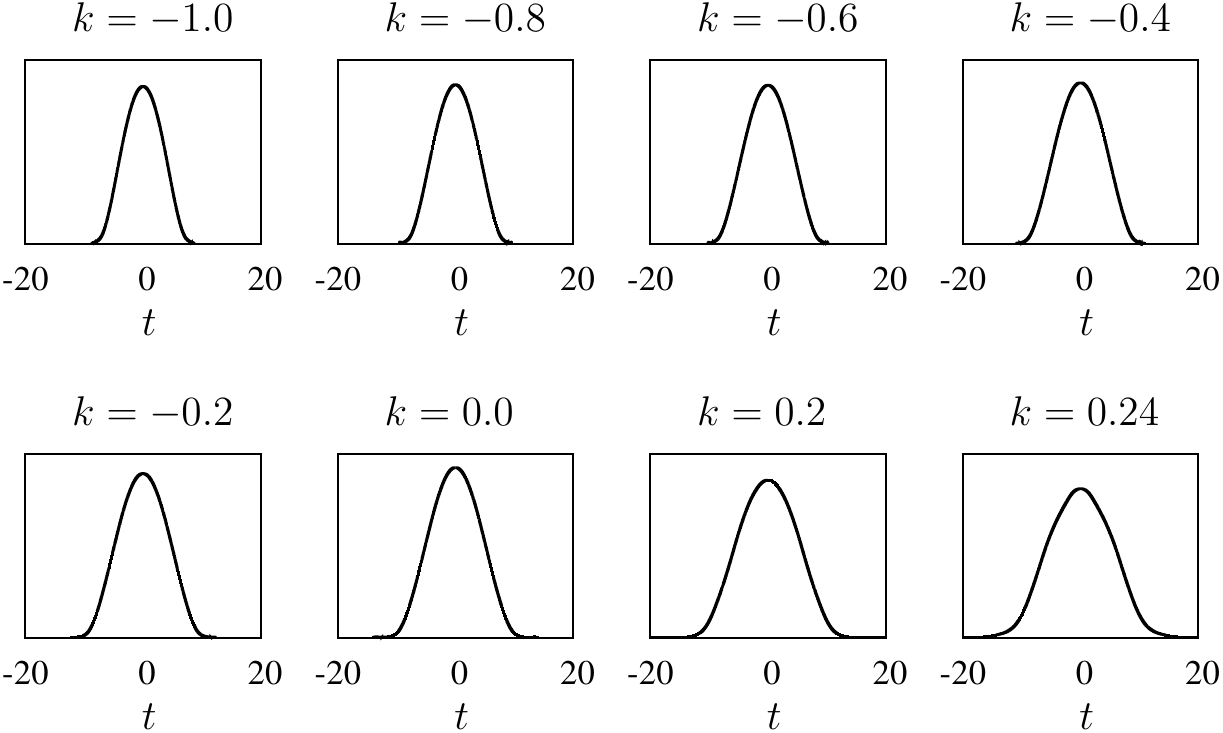}}}
\caption{Average volume profile $\langle N_2(t)\rangle$ as function of time $t$, 
measured at $\alpha\! =\! -1$ for various values of $k$. 
The scale of the vertical axis is the same for all plots. Some of the profiles have a tail which is not visible here, 
but shown in Fig.\ \ref{fig:vp_k_3} below.}
\label{fig:vp_k}
\end{figure}

In Sec.\ \ref{sec:numsetup} we introduced the notions of {\it time} and {\it spatial slices} for a general, nonfoliated CDT geometry.
The presence of these ingredients allows us to measure volume 
distributions -- also called {\it volume profiles} -- just like in standard CDT quantum gravity.
In the following we will present the results of our numerical investigations. 
Fig.\ \ref{fig:vp_k} shows the expectation value of the measured volume distributions for various 
values of the coupling $k$, for fixed $\alpha\! =\! -1$. 
In all cases the average geometry is macroscopically extended and the average volume profile has a
characteristic blob shape, strongly reminiscent of what is found in CDT in the physically interesting phase \cite{3dcdt}. 
We will report later in this section on a quantitative analysis of the average volume distributions.

\begin{figure}[b]
\centerline{\scalebox{1.0}{\includegraphics{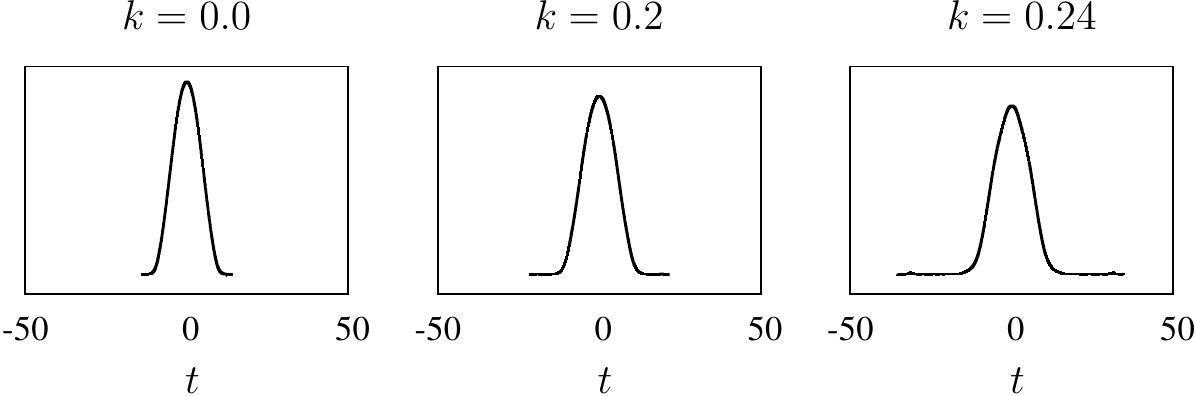}}}
\caption{The last three average volume profiles from Fig.\ \ref{fig:vp_k}, 
with a different scaling of the time axis and a small upward shift away from the axis to exhibit the tails of the distributions.}
\label{fig:vp_k_3}
\end{figure}
\begin{figure}[t]
\centerline{\scalebox{0.8}{\includegraphics{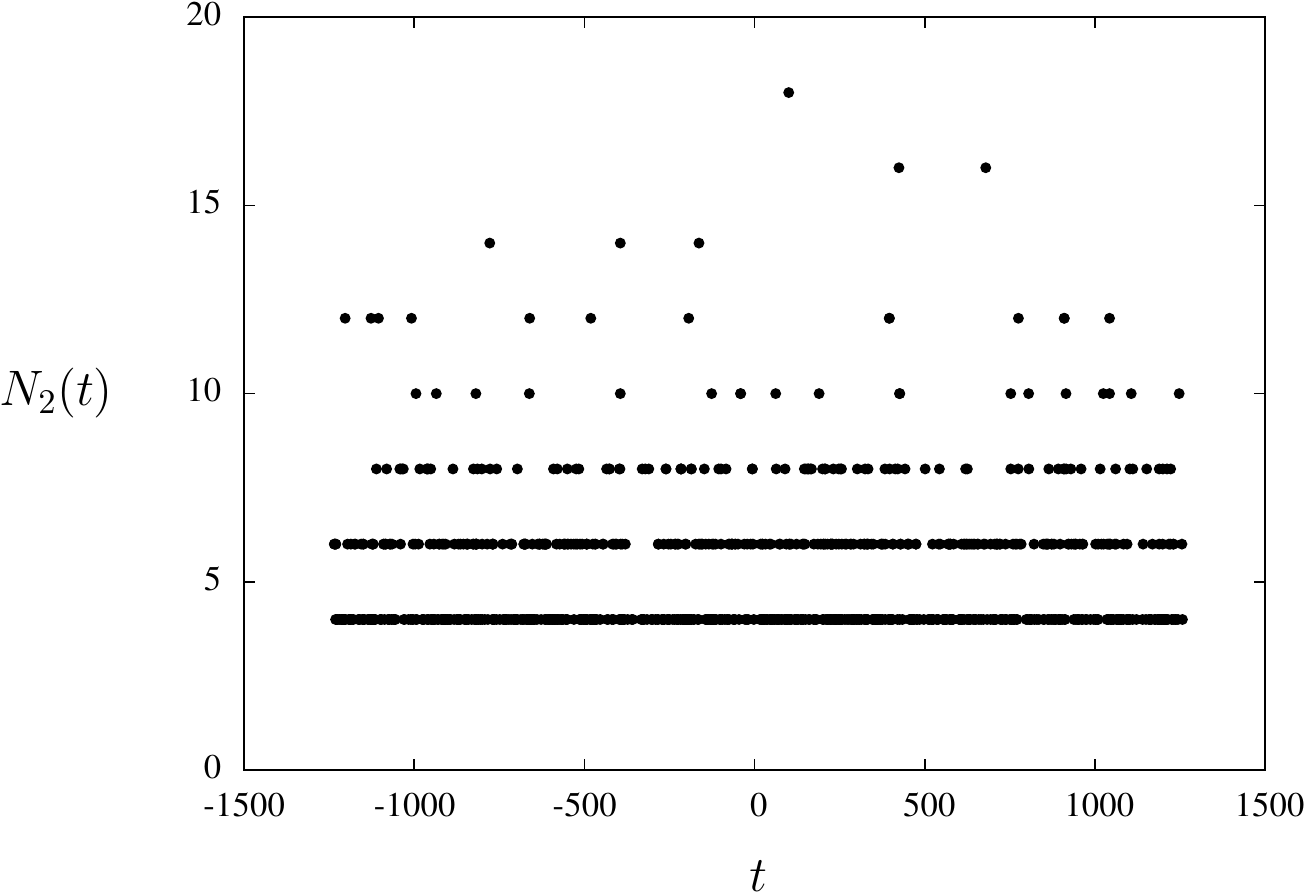}}}
\caption{Volume distribution $N_2(t)$ of a typical triangulation from a simulation at 
$(k,\alpha)\! =\! (0.4,-1)$ and $N_3=40.000$ in the phase of high vertex density. The plot shows
individual measurement points.}
\label{fig:vp_tube}
\end{figure}

Fig.\ \ref{fig:vp_k} illustrates that with increasing $k$ the time extension of the average geometry also increases. 
In addition, as one approaches the phase transition, the emergent geometry develops a ``tail" at both ends of the 
volume profile, by which we mean a region of small, approximately constant spatial volume. 
Since this structure is not resolved in Fig.\ \ref{fig:vp_k}, we have replotted 
the distributions close to the phase transition (at $k\! =\! 0,\, 0.2,\, 0.24$) in Fig.\ \ref{fig:vp_k_3}, with 
an enlarged scale for the time axis and a small upward shift of the distribution curves.
This tail looks similar to the stalk observed in simulations of CDT, but is not necessarily related because of the
different choices of boundary conditions. In CDT, its presence is enforced by the fact that the simulations are run
at a fixed total time extent (equal to the number of spatial slices in the foliation), {\it and} that the two-volume is not
allowed to vanish, but is bounded below by a minimum of four triangles, set by the manifold conditions.  
In the present case, we employ the same regularity condition, but 
the time extension of the geometry is dynamical and the stalk 
develops spontaneously as we move from $k\! =\! 0$ towards the phase transition. 
Anticipating our interpretation below of the volume profiles in terms of de Sitter universes, the
appearance of the tails could be related to quantum corrections to the underlying effective 
minisuperspace action near the phase transition.

Consider now a volume distribution on the line $\alpha\! =\! -1$ {\it beyond} the transition, that is,
in the phase of high vertex density.
Fig.\ \ref{fig:vp_tube} shows a volume distribution of a typical path integral configuration from a simulation at 
$(k,\alpha)\! =\! (0.4,-1)$ with 40.000 tetrahedra. 
The qualitative picture in this phase is completely different:  
the vast majority of spatial slices have (almost) minimal size $N_2(t)$, and the
triangulation forms a very long stalk with minimal spatial extent almost everywhere. 
At this phase space point, we have checked that the time extension of the stalk scales linearly with the system size. 
In the infinite-volume limit, it would therefore appear that the ``universe" becomes a one-dimensional timelike string.

We can now summarize our findings. At all phase space points investigated we have found average geometries 
which are macroscopically extended and whose volume profile has a characteristic blob-like shape.
The time extension of the average geometry increases with increasing $k$, and near the phase transition the
geometry starts to develop tails. On the other side of the transition the geometries degenerate into long 
tubes, unrelated to any 2+1 dimensional classical geometry.

\subsubsection*{Evidence for three-dimensionality from finite-size scaling}

We will investigate next whether we can assign a macroscopic dimensionality to the extended structure of the
volume profiles found in the phase of low vertex density, by performing a systematic finite-size scaling analysis.
To this end, we have run another extended series of simulations, taking data at six points along the axis of constant 
$\alpha\! =\! -1$, ranging from $k\! =\! -1.0$ to $k\! =\! 0.0$, and at six points along the axis of constant $k\! =\! 0$,
ranging from $\alpha\! =\! -0.52$ to $\alpha\! =\! -1.0$. At each point we have performed four simulations with different system 
sizes, namely, $N_3\! =\! 40$, 80, 120 and 160$k$.

\begin{figure}[t]
\centerline{\scalebox{1.2}{\includegraphics{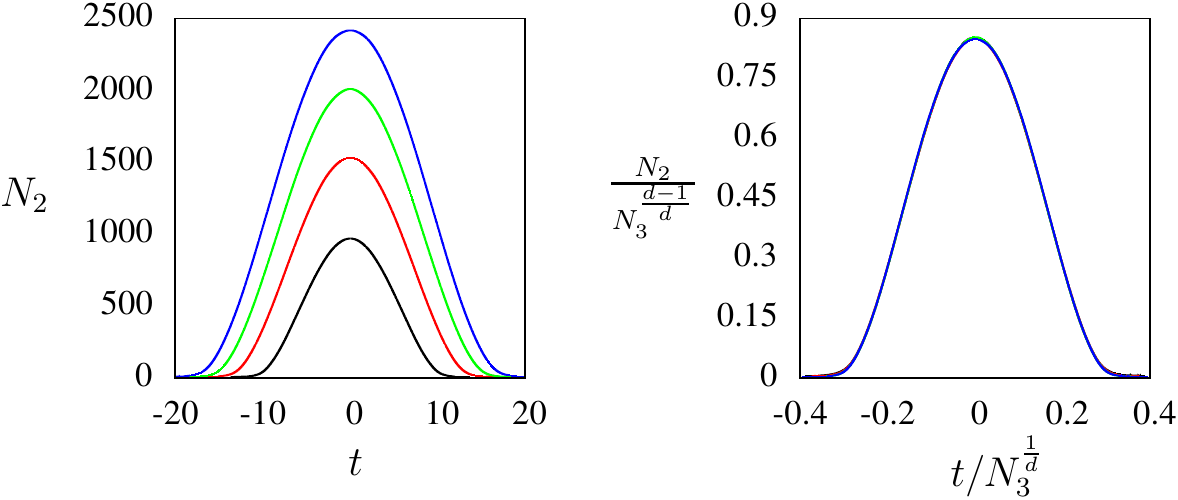}}}
\caption{Average volume profiles at $(k,\alpha)\! =\! (0.0,-1.0)$ for four different system sizes 
$N_3\! = $40, 80, 120 and 160$k$, before (left) and after (right) rescaling the axes as indicated, with $d\! =\! 2.91$ 
to achieve a best overlap, which is seen to be of excellent quality.}
\label{fig:vpfss_k0.0}
\end{figure}

Fig.\ \ref{fig:vpfss_k0.0} (left) shows the measurements of the expectation value of the volume 
distributions for the four different system sizes
at $(k,\alpha)\! =\! (0.0,-1.0)$. Following the strategy of \cite{desitter} for foliated CDT triangulations in 3+1 dimensions, 
we will use finite-size scaling to achieve a best overlap of these curves.
Assuming that the average geometry has macroscopic dimension $d$, we expect time intervals to scale like $N_3^{1/d}$ 
and spatial volumes like $N_3^{(d-1)/d}$. When plotting the distributions with axes rescaled accordingly, 
the measured curves should fall on top of each other.

To find an estimate for $d$ we have run an algorithm that scans through an interval of $d$-values in steps of 
$\Delta d=0.005$, which for each $d$-value measures how well the volume profiles overlap. 
We have employed a standard least-squares measure with appropriate normalization to quantify the quality of the overlap. 
The value of $d$ which minimizes this measure is taken as an estimate for the macroscopic dimension. 
For the case at hand, the algorithm yields a best estimate of $d\! =\! 2.91$. 
The plot in Fig.\ \ref{fig:vpfss_k0.0} (right) shows all four distributions with axes rescaled using this value 
for the dimension, resulting in a virtually perfect overlap.

\begin{figure}[t]
\centerline{\scalebox{1.2}{\includegraphics{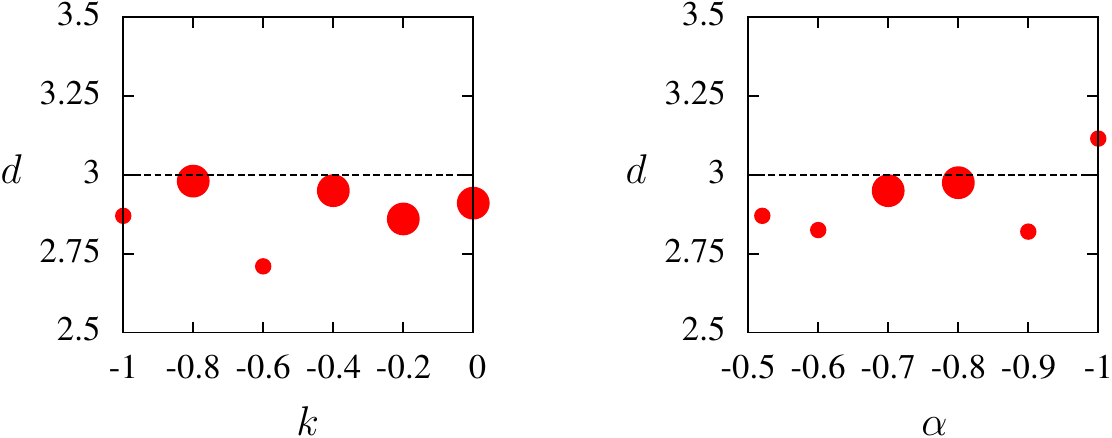}}}
\caption{Estimates for the macroscopic dimension $d$ from finite-size scaling, from measurements
at fixed $\alpha\! =\! -1.0$ (left plot) and fixed $k\! =\! 0.0$ (right plot). 
Large dots represent measurements where the volume profiles overlap with excellent quality, 
the smaller dots stand for overlaps of lesser quality.}
\label{fig:d}
\end{figure}

We have repeated the same analysis for the other points in the phase diagram. 
Fig.\ \ref{fig:d} summarizes the calculated estimates for the macroscopic dimension $d$, 
for fixed $\alpha\! =\! -1$ (left) and fixed $k\! =\! 0$ (right). The large dots indicate measurements with an 
overlap of excellent quality, the small dots those of a somewhat lesser quality.
We observe that all six high-quality measurements yield macroscopic dimensions between $d\! =\! 2.85$ and $d\! =\! 3.00$, 
while the values from the remaining six measurements have a larger spread.

We have not included error bars in the plots of Fig.\ \ref{fig:d} and the values obtained for the dimension $d$,
because they are dominated by systematic errors
we currently cannot estimate, one possible source being algorithmic dependences. 
The calculation of the dimension observable is highly nontrivial and involves several algorithmic choices
which potentially affect the final result. Recall that we first had to define a time coordinate, which we did by calculating the 
average distance between a vertex and the poles of the three-sphere. Secondly, we assigned a time coordinate to the 
spatial slices, by averaging over the time coordinates of the vertices in the slice. 
Finally, we ran a rather sophisticated averaging algorithm to produce the final distributions. 
The nonuniqueness of this entire process is likely to lead to systematic errors not captured by standard error 
algorithms such as the bootstrap method.

It is clear from this discussion that a single dimension measurement does not provide sufficient evidence 
to support the $d\! =\! 3$ hypothesis, which would imply compatibility with the standard CDT result.
On the other hand, all twelve results together suggest strongly that the average geometries in the 
phase of low vertex density are three-dimensional. The results on 
the functional form of the volume profiles presented below will strengthen this preliminary conclusion
even further.

\subsubsection*{Comparison with the three-sphere}

\begin{figure}[t]
\centerline{\scalebox{0.9}{\includegraphics{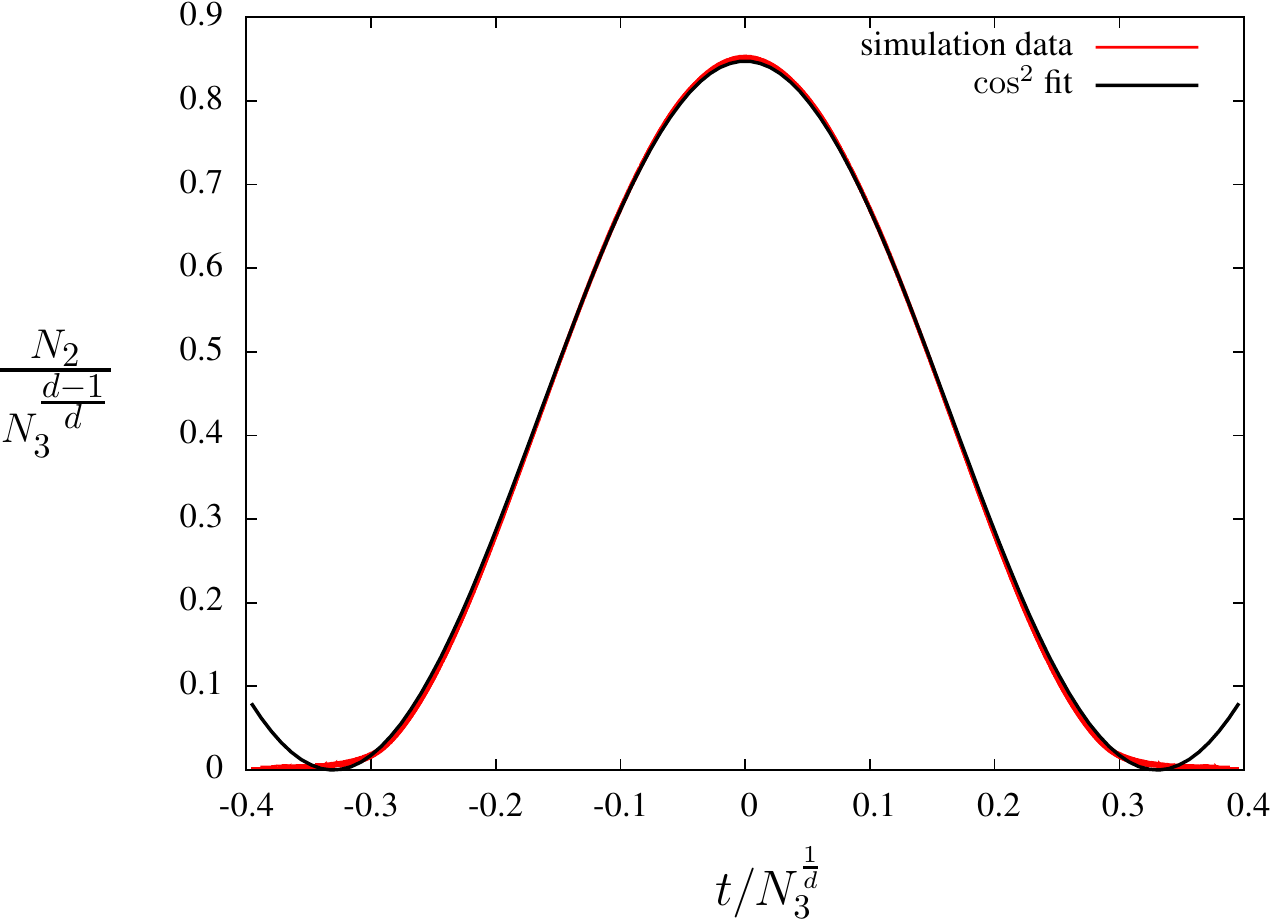}}}
\caption{Rescaled average volume profiles at $(k,\alpha)\! =\! (0.0,-1.0)$,
corresponding to dimension $d\! =\! 2.91$, and best fit to the $\cos^2$-ansatz.}
\label{fig:fit_k0.0}
\end{figure}

A crucial piece of evidence that CDT quantum gravity has a well-defined classical limit comes from
matching the average distributions of spatial volumes with those of a Wick-rotated version of a
solution to the classical Einstein equations, namely, a de Sitter universe \cite{desitter}. 
More specifically, the distributions coming from the simulations have been compared 
to a volume profile of the form $V_3(t)=a \cos^3(b t)$, where $t$ is by assumption proportional to
proper time. This is the volume profile of Euclidean de Sitter space (equivalently, the round four-sphere), 
where the two free parameters $a$ and $b$ depend on the overall size of the universe and a 
finite relative scaling between spacelike and timelike directions. The measured volume profiles
in 3+1 dimensional CDT can be fitted with high accuracy to the analytical $\cos^3$-expression \cite{desitter,semicl}, 
with the exception of the region very close to the end points of the curve, which cannot be resolved with
sufficient precision and is obscured by the regularity condition $\langle N_2(t)\rangle\! \geq\! 4$, as we
have discussed earlier.

\begin{figure}[t]
\centerline{\scalebox{0.9}{\includegraphics{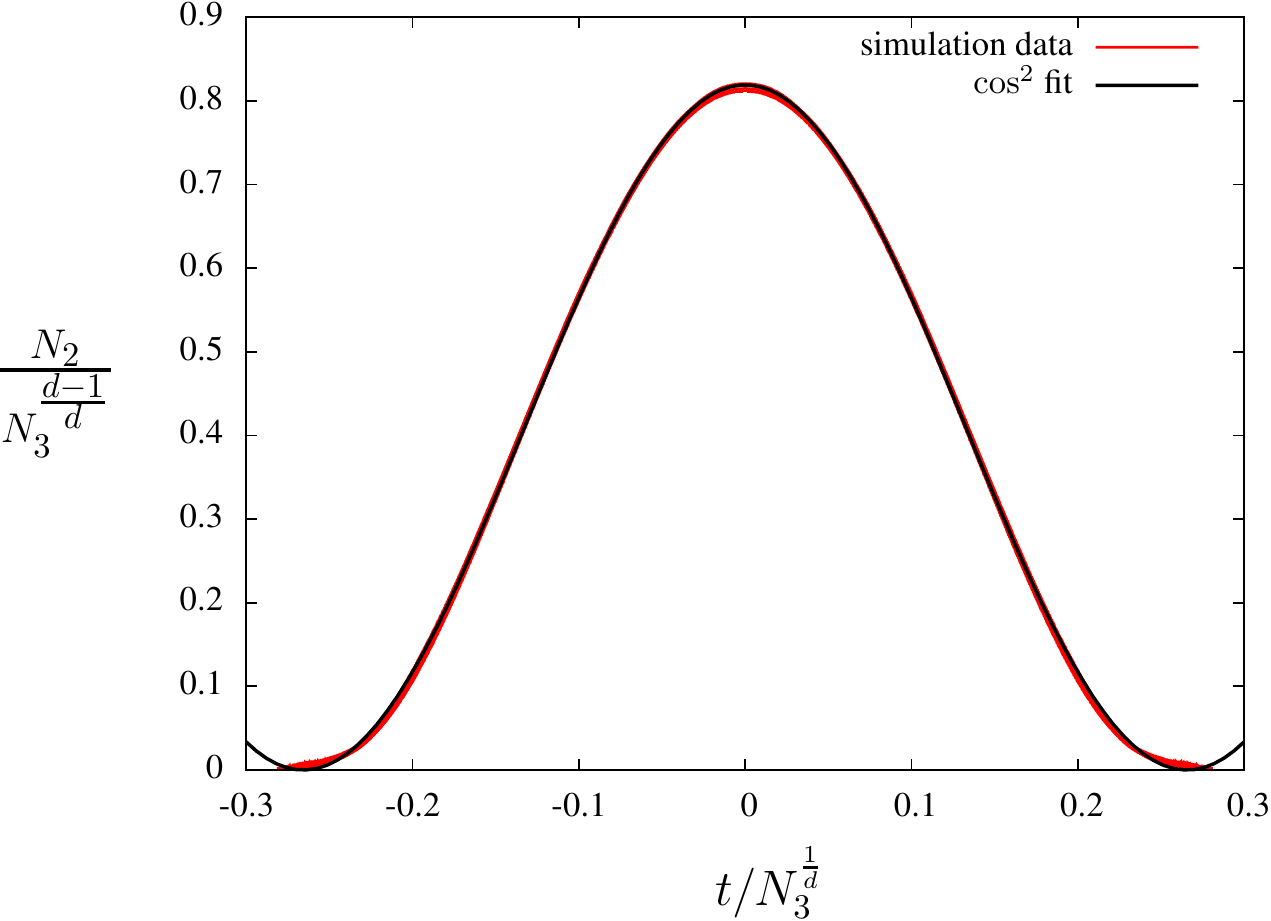}}}
\caption{Rescaled average volume profiles at $(k,\alpha)\! =\! (-0.8,-1.0)$,
corresponding to dimension $d\! =\! 2.98$, and best fit to the $\cos^2$-ansatz.}
\label{fig:fit_k-0.8}
\end{figure}

We will perform an analogous analysis of nonfoliated CDT in 2+1 dimensions, using
the average volume distributions from our simulations. 
The volume profile of the corresponding continuum de Sitter universe in three dimensions has the 
functional form $V_2(t)\! =\! a \cos^2(b t)$, where $a$ and $b$ are constants. To extract an optimal fit to this 
two-parameter family of curves from our Monte Carlo data, we have selected only those points in the phase diagram 
where the rescaled average volume profiles overlap with excellent quality, and where we have a well-defined curve
to compare to.

Fig.\ \ref{fig:fit_k0.0} shows the outcome of this comparison at the point $(k,\alpha)\! =\! (0.0,-1.0)$. 
Obviously, the only relevant part of the fit function is the region between the two zeros of the $\cos^2$-function.
We see that the functional ansatz fits the average volume distributions almost perfectly, 
except at the two ends, where the simulation data show a small tail which is not present in the fit function 
once we cut away the parts outside the two minima. 
We have already commented earlier on the appearance of such tails in the vicinity of the phase
transition (see also Fig.\ \ref{fig:vp_k_3}); in the context of the de Sitter interpretation of our universe
they indeed seem to be related to small-scale deviations from the classically expected result.

From this point of view it is interesting to understand how the situation changes when one repeats the
comparison at a point further away from the phase transition. 
Fig.\ \ref{fig:fit_k-0.8} shows the result of the same analysis performed at the point 
$(k,\alpha)\! =\! (-0.8,-1.0)$. We again observe an almost almost perfect fit in the region where the spatial volume
$N_2(t)$ is nonminimal. Remarkably, now even the total time extension of the dynamically generated
universe and the de Sitter fit function between the two minima agrees, and we get an almost perfect
semiclassical matching. The quality of the fit becomes slightly reduced 
towards both ends, which is not surprising because discretization effects become large when the 
spatial volumes become small.

\section{Summary and conclusions}
\label{sec:conclusions}

We begun our investigation with the aim to isolate and understand the role of the preferred time slicing in standard
CDT quantum gravity, while maintaining causality of the individual path integral histories. 
In this article we have presented many details of the kinematical and dynamical properties of the new,
``nonfoliated" CDT model in 2+1 dimensions, which implements the dissociation of the causal 
structure and the preferred notion of time. Due to the presence of new elementary building blocks, the foliation
in terms of equally spaced triangulated spatial hypermanifolds is broken up in this extended version of CDT,
acquiring novel simplicial substructures such as ``bubbles" and ``pinchings".
 
Gravitational dynamics in the new model is implemented in terms of the standard Regge action,
defined as a linear function on the space of independent counting variables, which is five-dimensional, compared
to CDT's two dimensions.
After fixing the total system size, there are two coupling constants spanning the phase space of the model,
the bare inverse Newton coupling $k$ and the coupling $\alpha$, which quantifies the anisotropy between space- and timelike 
length assignments in the regularized theory. This asymmetry parameter has to satisfy the inequalities 
$1/2\! <\! | \alpha | \! <\!  3$ for the Wick rotation to exist, which is necessary to be able to probe the nonperturbative
properties of the model with the help of Monte Carlo simulations. This introduces two boundaries in the phase diagram.

The presence of thermalization problems, preventing the effective implementation of the Monte Carlo algorithm,
led us to eliminate certain global simplicial substructures.
This allowed us to investigate the region $-1\!\lesssim\! \alpha\! <\! -1/2$ of the phase diagram (in terms
of the analytically continued $\alpha$), while we still observed severe thermalization problems in the complementary
region. We ran the simulations with the spacetime topology of a three-sphere with a source and a sink of (Euclidean) 
time at the two poles.

In terms of results, we have found two phases of geometry with low and high vertex density, for $k$-values below 
and above some critical value $k_c$ of the inverse gravitational coupling respectively. 
The analysis of the tetrahedron distributions revealed that the triangulations remain weakly foliated throughout
the investigated phase space region of low vertex density, but that the strength of this signal varies significantly
as function of the bare couplings.
In addition, we observed the emergence of almost perfectly 
foliated simplicial geometries close to the boundary $\alpha\! =\! -1/2$ of the phase diagram. 
We constructed a volume distribution observable and an averaging procedure to study the expectation value of 
the volume profiles of the emergent geometries in the weakly foliated phase. 
A finite-size scaling analysis provided strong evidence that the extended geometries are
macroscopically three-dimensional. Additional support for this came from fitting the measured profiles to a 
$\cos^2$-ansatz corresponding to a classical de Sitter universe, which found an almost perfect agreement. 
We have repeated the analysis for various points in the phase diagram, giving consistent results.

These results provide compelling evidence that the phases of low vertex density of both foliated and 
nonfoliated CDT quantum gravity have the same large-scale properties in the continuum limit and
lie in the same universality class. Since apart from removing the {\it distinguished} time slicing we  
essentially left all other ingredients of the kinematics intact, this would imply that the presence or absence of
a preferred foliation in CDT is not a relevant ingredient.
As remarked already in the introduction,
the same is not true for Ho\v rava-Lifshitz gravity \cite{hl} -- to the extent that our nonperturbative, 
coordinate-free set-up can be
compared with this continuum formulation -- where a fixed spatial foliation is essential. 
It does not mean that CDT, or suitable extensions like that studied in \cite{cdthorava}, cannot provide a 
framework suitable for studying anisotropic gravity models. 
Our results also conform with the expectation that in 2+1 dimensions the value of the parameter $\alpha$
is irrelevant from the point of view of the continuum theory.

Because of the strong similarities of the large-scale
properties of CDT quantum gravity in 2+1 and 3+1 dimensions, it is plausible to conjecture that also in
3+1 dimensions the presence or absence of a 
direct-product structure of the triangulations does not influence the final outcome.
If this is the case, one may want to stick with the simpler ``standard" formulation of Causal Dynamical Triangulations
as a matter of convenience and computational simplicity, as we have already pointed out elsewhere \cite{jl}.

\vspace{.5cm}

\noindent {\bf Acknowledgements.} 
We thank J. Ambj\o rn for a critical reading of the manu\-script.
The authors' contributions are part of the research programme of the Foundation for Fundamental Research 
on Matter (FOM), financially supported by the Netherlands Organisation for Scientific Research (NWO).
The work was also sponsored by NWO Exacte Wetenschappen (Physical Sciences) for the use 
of supercomputer facilities, with financial support from NWO.

\end{document}